\documentclass[reprint,
 amsmath,amssymb,
 aps,
pra,
floatfix]{revtex4-2}

\usepackage{outlines}
\usepackage{dsfont}
\usepackage{enumitem}
\usepackage{braket}
\usepackage{amsmath}
\usepackage{mathtools}
\usepackage{amsfonts,amssymb,amsbsy}
\usepackage{graphicx}
\usepackage{MnSymbol}
\usepackage{xcolor}
\usepackage{hyperref}
\usepackage{physics}
\usepackage{array}
\usepackage{multirow}

\newcommand{\<}{\langle}

\renewcommand{\>}{\rangle}
\renewcommand{\(}{\left(}
\renewcommand{\)}{\right)}
\renewcommand{\[}{\left[}
\renewcommand{\]}{\right]}

\newcommand{\OO}{\mathcal{O}}

\newcommand{\C}{\mathcal{C}}
\newcommand{\HH}{\mathcal{H}}
\newcommand{\LL}{\mathcal{L}}

\hypersetup{
    colorlinks=true,
    linkcolor=blue,
    filecolor=blue,      
    urlcolor=blue,
    citecolor=blue
}

\begin{document}


\title{Randomized Benchmarking with Leakage Errors}
\author{Yi-Hsiang Chen}
\email{yihsiang.chen@quantinuum.com}
\affiliation{Quantinuum, 303 South Technology Court, Broomfield, CO 80021, USA}

\author{Charles H. Baldwin}
\email{charles.baldwin@quantinuum.com}
\affiliation{Quantinuum, 303 South Technology Court, Broomfield, CO 80021, USA}

\date{\today}

\begin{abstract}
Leakage errors are unwanted transfer of population outside of a defined computational subspace and they occur in almost every platform for quantum computing. While prevalent, leakage is often overlooked when measuring and reporting the fidelity of quantum gates with standard methods. In fact, when leakage is substantial it can cause a large overestimation of fidelity from the typical method used to measure fidelity, randomized benchmarking. We provide several methods for properly estimating fidelity in the presence of leakage errors that are applicable in different error regimes with carefully chosen sequence lengths.  Then, we numerically demonstrate the methods for two-qubit randomized benchmarking, which often have the largest errors. Finally, we reanalyze previously shared data from Quantinuum systems with some of the methods provided.
\end{abstract}

\maketitle


\section{Introduction}
Quantum computer performance is currently limited by errors, which occur in all components of quantum circuits. Errors come in a variety of flavors but a particularly nefarious and often overlooked type is leakage errors. Roughly, leakage errors move population from the desired computational states into other ``leaked'' states. Leakage errors exist in all varieties of quantum computing systems, for example atoms (ions~\cite{Ozeri07,Moore23} and neutrals~\cite{Evered23}) with decay to undesired atomic sublevels or even atom loss, superconductors~\cite{Wood18,Acharya24} with unwanted coupling to higher non-harmonic levels, or silicon quantum dots~\cite{Andrews19} with different permutations of electron states. Leakage errors are often due to a fundamental constraint of the system's design, e.g. spontaneous emission~\cite{Ozeri07,Moore23} with laser-based gates on atoms, and are particularly detrimental to near term applications, like Hamiltonian simulation~\cite{Chertkov24}, and longer-term in fault-tolerant quantum computing with quantum error correction~\cite{Brown20, Acharya24}. Despite the prevalence of leakage errors and their importance in near- and long-term quantum computing, they are often not measured and reported especially for two-qubit (2Q) gates. 

\begin{figure} 
\centering
\includegraphics[width=\columnwidth]{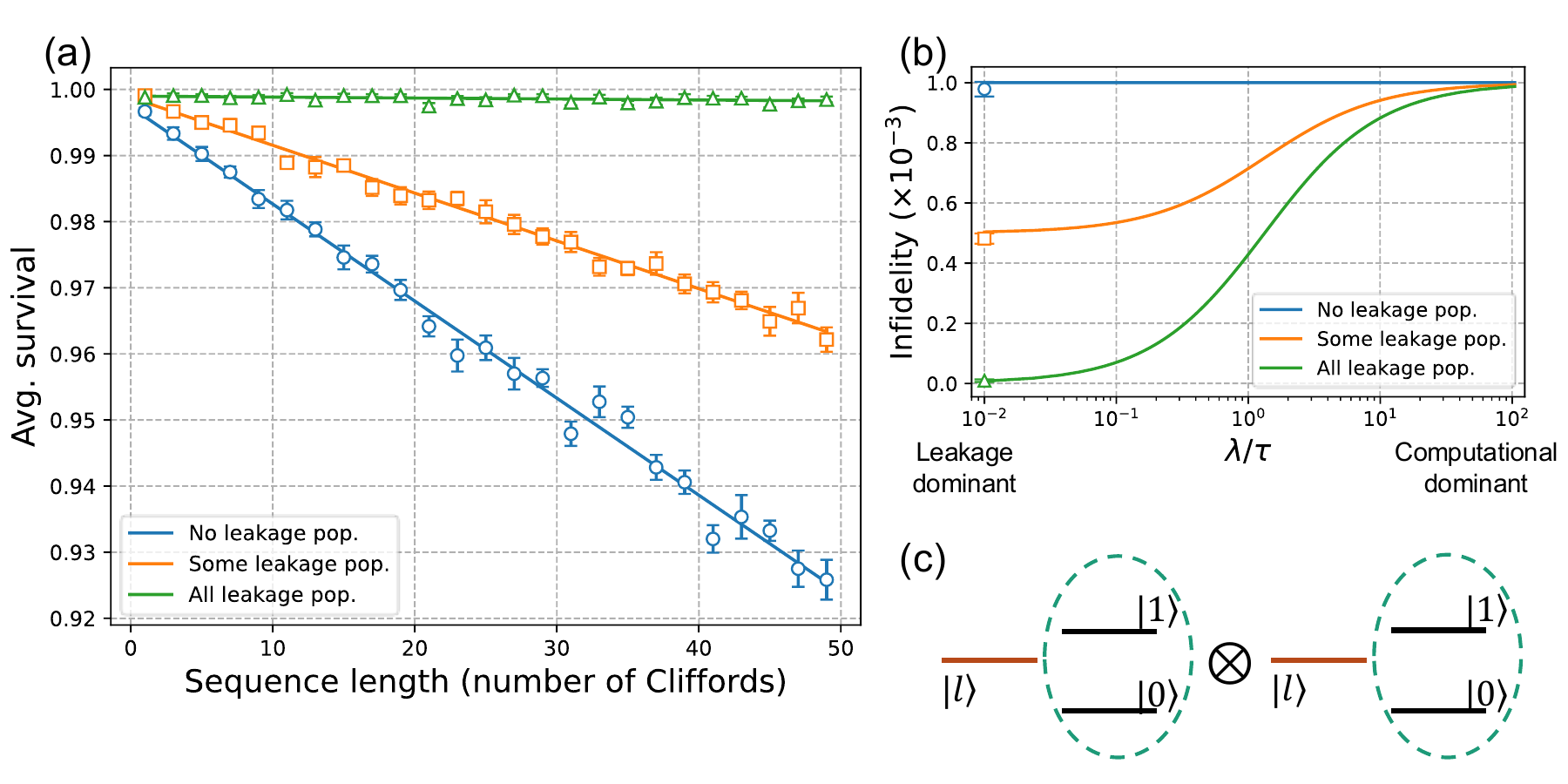}
\caption{Clifford 2Q RB with leakage returns different fidelity estimates with measurement operators that contain different amounts of projection onto leakage subspaces. (a) An example process with a fixed infidelity $10^{-3}$ that is dominated by leakage as described in Sec.~\ref{sec:when_leakage_ignored}. Clifford 2Q RB survival probability plot for three different measurement settings, from a measurement operator with no leakage population, some leakage population to all leakage population. (b) Infidelity comparisons between three different methods for the example process with a changing leakage magnitude but a fixed total infidelity of $10^{-3}$. The quantity $\lambda$ represents the magnitude of computational errors and $\tau$ represents the magnitude of leakage errors. (c) Energy level diagram of two qubits each with a leakage state $\ket{l}$.\label{fig:leakage_RB_comparisons} }
\end{figure}

Quantum computer performance is typically measured with randomized benchmarking (RB)~\cite{Magesan12}. RB has many variants to quantify and characterize various errors, including leakage. In most applications of RB, the goal is to quantify the overall gate performance called fidelity~\cite{Andrews19, Moses23, McKay23, Evered23, Acharya24}. It is generally believed that different variants of RB that are designed to measure fidelity generally return similar results~\cite{Polloreno23} (perhaps up to small multiplicative factors). However, when leakage errors are present, this may no longer be the case and different choices of the measurement operator may lead to drastically different fidelity estimates. For example, Fig.~\ref{fig:leakage_RB_comparisons}a plots a simulation of standard 2Q Clifford RB~\cite{Magesan12} with an error process that is dominated by leakage errors (far left of Fig.~\ref{fig:leakage_RB_comparisons}b). The RB experiment is only run for short sequence lengths (linear decay) with different measurement operators. It is typically assumed that the slope of the decay is directly related to gate fidelity, but since the slope depends on the measurement operator this relation can be broken with leakage errors. Therefore, standard RB results can diverge with leakage errors and in many cases overestimate performance. Additionally, with longer sequence lengths the decay functions become more complicated and do not necessarily have a single exponential term, as is typically assumed. This can cause more issues with fitting and even in the best case (Fig.~\ref{fig:leakage_RB_comparisons}b blue line, measuring no leakage population) may lead to overestimates of fidelity.

There are many proposed methods to measure the fidelity in the presence of leakage~\cite{Epstein14,Andrews19,Chasseur15,Wallman15,Wallman16,Wood18,Wu24}. The effect of leakage in randomized benchmarking was originally studied in Ref.~\cite{Epstein14} and shown to complicate the standard fitting used in RB of single-qubit (1Q) gates. Ref.~\cite{Epstein14} also proposed a modification to standard RB that required additional control to randomize the phase between computational and leakage subspaces to reduce the complexity of the fitting and extract leakage rates. Ref.~\cite{Chasseur15} generalized that approach to measure the fidelity with leakage, where parameters are obtained by fitting from a multi-exponential decay but the method could not extract leakage rates.

Independently, Ref.~\cite{Wallman15} proposed an RB variant to measure loss (leakage without the possibility for return to the computational subspace) for arbitrary number of qubits. Later, Refs.~\cite{Wallman16,Wood18} proposed methods to extract leakage rates from 1Q experiments that again required phase randomization between computational and leakage states but extracted the desired rates and Ref.~\cite{Wood18} showed how those rates relate to 1Q computational fidelity. Ref.~\cite{Claes21} showed how similar methods applied for 1Q gates even without phase randomization when the leakage operations are known. Ref.~\cite{Wu24} extended this work to multiple qubits with similar phase randomization constraints in addition to new assumptions about the leakage process. Leakage has also been measured in different systems before. Many either use potentially platform-dependent control ability to measure computational and leakage errors separately for 1Q gates~\cite{Chen16} or use heuristic modeling of the error process~\cite{Andrews19,Radnaev24,Muniz24} with post-selection~\cite{Radnaev24,Muniz24} to obtain the leakage rate and the fidelity.

In this paper, we propose new and efficient RB methods that account for leakage in \emph{multi-qubit} gates. We introduce three main methods in Sec.~\ref{sec:rb_methods} and discuss four different error regimes where the fidelity and the leakage rate can be reliably obtained. We also show evidence that not properly accounting for leakage errors has led to small underestimation of 2Q gate errors in previous Quantinuum system data. While the impact is barely above statistical noise, it is possible that similar choices in the future could lead to bigger discrepancies. 

This paper is organized as follows. We start with a general description of the leakage process in RB and the definition of averaged fidelity in Sec \ref{sec:main_result}. We then propose RB methods in Sec~\ref{sec:rb_methods} and derive survival probabilities that inform error parameters in Sec.~\ref{sec:survival_probs}. Numerical tests are also provided in Sec.~\ref{sec:survival_probs} that verify our methods effectiveness. Finally, in Sec.~\ref{sec:machine_data}, we re-analyze the existing Quantinuum machine data using the methods provided.

\section{Leakage processes}\label{sec:main_result}
Leakage errors move population from the computational subspace to other ``leaked'' states. Most gate-based quantum computing uses qubits (two-dimensional Hilbert spaces) that are embedded in larger Hilbert spaces (e.g.~\ref{fig:leakage_RB_comparisons}c). There are often processes that couple qubit states to leaked states at some rate. For example, a trapped-ion hyperfine qubit occupies two magnetic sublevels in the ground-state manifold of an atomic ion species but, depending on the nuclear spin of the ion, there might be additional magnetic sublevels that correspond to leaked states. The other sublevels might couple to the qubit subspace by stray magnetic fields or, most likely, spontaneous emission from scattering photons off excited states used for laser-based gates.

In order to extract fidelity from RB when leakage errors are present, we must model the effect of leakage errors on standard RB sequences. For now, we ignore SPAM errors for simplicity, but consider their effects in App.~\ref{app:SPAM_surv}. 

\subsection{Leakage subspace}
To begin, we decompose the total Hilbert space $\HH$ into the computational space $\HH_C$ with dimension $d_C$ and the leakage space $\HH_L$ with dimension $d_L$, i.e. $\HH=\HH_C\oplus \HH_L$. Denote the basis in the computational space $\{|i\rangle\}_{\HH_C}$ and the basis in the leakage space $\{|\alpha\rangle\}_{\HH_L}$. Any operator $\rho\in\LL(\HH)$ on the Hilbert space can also be decomposed into computational and leakage components by $\chi_C\oplus \chi_L$, where we define
\begin{align}
&\chi_C=\text{Span}\{|i\rangle\langle j|\}_{i,j\in\HH_C} \nonumber\\
&\chi_L=\text{Span}\{|i\rangle\langle\alpha|,|\alpha\rangle\langle i| \}_{i\in\HH_C,\alpha\in\HH_L} \oplus \text{Span}\{|\alpha\rangle\langle \beta|\}_{\alpha,\beta\in\HH_L}. \nonumber
\end{align}
An operator $\rho$ can then be written as a direct sum of the component in each subspace $\rho_C\in\chi_C$ and $\rho_L\in\chi_L$, i.e.,
\begin{align}
\rho&= \rho_C \oplus \rho_L =\underbrace{\sum_{i,j\in\HH_C} \rho_{ij} |i\rangle\langle j|}_{\rho_C} \nonumber\\
&+ \underbrace{\sum_{i\in\HH_C,\alpha\in\HH_L}\left(\rho_{i\alpha} |i\rangle\langle\alpha|+\rho_{\alpha i} |\alpha\rangle\langle i|\right) +\sum_{\alpha,\beta\in\HH_L} \rho_{\alpha\beta} |\alpha\rangle\langle \beta|}_{\rho_L}.
\end{align}

Processes map operators to operators $\mathcal{B}(\mathcal{H}): \LL(\HH) \rightarrow \LL(\HH)$. Define an error process $\Lambda$ that may map parts of the computational subspace to the leakage subspace as
\begin{align}
\Lambda=\begin{pmatrix}
\Lambda_{CC} & \Lambda_{CL} \\
\Lambda_{LC} & \Lambda_{LL}
\end{pmatrix},
\end{align}
where $\Lambda_{LC}$ represents leakage, i.e., terms that map computational to leakage subspace and $\Lambda_{CL}$ represents seepage, i.e., terms that map leakage to computational subspace. If the initial state is prepared in the computational subspace then seepage only happens after a leakage process, and therefore if the rates are roughly similar then seepage is a second-order effect \footnote[1]{More explicitly, let us consider an error process with a leakage error probability of $p_L$ and a seepage probability $p_S$. For a circuit with $\ell$ gates, the probability of one leakage event is $p(\textrm{one leakage})=\ell p_L(1-p_L)^{\ell-1}=\OO(\ell p_L)$. If the quantum state starts in the computational space, then a seepage event can happen only after a leakage event has happened. Therefore, the probability of a one seepage event is $p(\textrm{one seepage})=\sum_{k=1}^\ell(\ell-k)p_S(1-p_S)^{\ell-k-1}p_L(1-p_L)^{\ell-1}=\OO(\ell^2p_Lp_S)$, which is a second-order effect when $p_S\approx p_L$.}.

Ideal gates are assumed unitary and represented by processes that do not have leakage errors. Seepage is forbidden due to the unitarity of the gate on the total Hilbert space $\HH=\HH_C\oplus\HH_L$. App.~\ref{app:block-diagonal_form} shows that a noiseless ideal gate $\C$ on the joint space $\chi_C\oplus\chi_L$ is block diagonal in the defined basis
\begin{equation}
    \C=\begin{pmatrix}
    \C_C & 0\\
    0 & \C_L
    \end{pmatrix}
\end{equation}
where $\C_C$ and $\C_L$ are unitary channels in the respected subspaces.

\subsection{Randomized benchmarking sequences with leakage}
To perform RB, a series of $\ell$ random gates are applied to an initial state $\rho_{\textrm{in}}$ followed by a final inverse gate that ideally undoes all previous evolution, and the result is measured with projector $\Pi_{\textrm{out}}$. Without errors, the initial state and measurement are typically selected such that $\Tr\left[\Pi_{\textrm{out}} \rho_{\textrm{in}}\right]=1$. There are several options for random gate selection but here we focus on the standard randomization over the Clifford group~\cite{Magesan12}. When errors are present the final inverse gate will not perfectly undo all previous evolution and the overlap between the final state and the measurement is not one. We call this overlap the survival probability and measure its decay as a function of sequence length $\ell$. 

Since the goal is to do quantum computation in the computational state it is safe to assume that $\rho_{\textrm{in}} \in \chi_C$. However, since $\Pi_{\textrm{out}}$ is a measurement operator it must form a complete positive operator-valued measure (POVM) with a set of other measurement operators $\{\Pi_k\}$ where $\sum_k \Pi_k = \mathds{I}$, which is the identity on the combined Hilbert space $\HH$. This implies that some elements of the POVM have projection onto the leakage subspace. The exact overlap will depend on the physical implementation.

Proceeding with the standard RB derivation, we average over all Clifford sequences, which acts to ``twirl'' the error process $\Lambda$ of each noisy gate into $\bar{\Lambda}$, i.e.,
\begin{align}\label{eq:lambda_bar}
&\bar{\Lambda}= \frac{1}{|\bold{C}|}\sum_{\C\in\bold{C}}\C^{-1}\Lambda\C \nonumber\\
&= \begin{pmatrix}
\underbrace{\frac{1}{|\bold{C}|}\sum_{\C\in\bold{C}}\C_C^{-1} \Lambda_{CC} \C_C}_{:=\bar{\Lambda}_{CC}}  & \underbrace{\frac{1}{|\bold{C}|}\sum_{\C\in\bold{C}}\C_C^{-1} \Lambda_{CL} \C_L}_{:=\bar{\Lambda}_{CL}}\\
\underbrace{\frac{1}{|\bold{C}|}\sum_{\C\in\bold{C}}\C_L^{-1} \Lambda_{LC} \C_C}_{:=\bar{\Lambda}_{LC}}  & \underbrace{\frac{1}{|\bold{C}|}\sum_{\C\in\bold{C}} \C_L^{-1} \Lambda_{LL} \C_L}_{:=\bar{\Lambda}_{LL}}
\end{pmatrix}.
\end{align}
The computational-to-computational map $\bar{\Lambda}_{CC}$, i.e., the upper-left block in Eq.~\eqref{eq:lambda_bar}, becomes a non-trace-preserving depolarizing process
\begin{equation} \label{eq:comp_depolarizing}
    \bar{\Lambda}_{CC}(\cdot)=r \mathcal{I}_{C}(\cdot)+\lambda\Tr\[\mathcal{I}_C(\cdot)\]\tfrac{\mathds{I}_C}{d_C},
\end{equation}
where $\mathcal{I}_C(\rho)=\rho_C$ projects out the computational component $\rho_C\in\chi_C$ of an input operator $\rho$. We define $r$ as the depolarizing parameter and $\lambda$ as the computational error. Furthermore, we define 
\begin{align}
    t:= \textrm{Tr}\left[ \mathds{I}_C \Lambda\(\mathds{I}_C/d_C\) \right] \label{eq:t}
\end{align}
as the computational population, which is related to the leakage rate $\tau$ defined in \cite{Wood18} as $t=1-\tau$. From the trace-preserving property of $\bar{\Lambda}$, we have 
\begin{align}
&\Tr\[\bar{\Lambda}(\mathds{I}_C/d_C)\]=1\nonumber\\
&=\Tr\[\mathds{I}_C\bar{\Lambda}_{CC}(\mathds{I}_C/d_C)\]+\Tr\[\mathds{I}_L\bar{\Lambda}_{LC}(\mathds{I}_C/d_C)\]\nonumber\\
&\implies 1=r+\lambda +\Tr\[\mathds{I}_L\Lambda(\mathds{I}_C/d_C)\] \nonumber\\
&\implies \lambda=t-r. \label{eq:lambda}
\end{align}
The computational block of the twirled channel becomes
\begin{equation} \label{eq:comp_depolarizing_r_t}
    \bar{\Lambda}_{CC}(\cdot)=r \mathcal{I}_{C}(\cdot)+(t-r)\Tr\[\mathcal{I}_C(\cdot)\]\mathds{I}_C/d_C
\end{equation}
in terms of the depolarizing parameter $r$ and the computational population $t$, where $r=t=1$ when there is no error. Equivalently, one can rewrite the channel $\bar{\Lambda}_{CC}$ in terms of the leakage rate $\tau$ and the computational error $\lambda$ by replacing $r\to 1-\lambda-\tau$ and $(t-r)\to\lambda$. In the next section, we find that it can be more convenient to use either $\(r,t\)$ or $\(\lambda,\tau\)$ depending on the cases.

After running a length $\ell$ random RB sequence and applying a final inversion gate the probability of observing the expected output is
\begin{equation} \label{eq:general_survival}
    p(\ell) = \Tr\left[ \Pi_{\textrm{out}} \bar{\Lambda}^{\ell}\left(\rho_{\textrm{in}}\right) \right],
\end{equation}
which we refer to as the survival probability. To extract fidelity from RB the common approach is to show that a general survival probability function has parameters that relate to fidelity, without many assumptions on the error present. Then measuring the survival probability for many values of $\ell$ and fitting the data to the expected survival probability decay provides the estimation of the fidelity. In general, it is difficult to solve for $\bar{\Lambda}^{\ell}$ exactly in the presence of leakage and therefore difficult to write a general survival probability from Eq.~\eqref{eq:general_survival} to fit the data and estimate the fidelity. However, we will show that with carefully chosen sequence lengths $\ell$ for different error regimes, it is possible to write down expressions for $\bar{\Lambda}^{\ell}$, and therefore $p(\ell)$.

\subsection{Leakage errors and fidelity} \label{sec:fidelity}

One may be tempted to say that computational errors affect $r$ and leakage errors affect $t$ but that is not the full story. In fact, $r \leq t$ \footnote[2]{To show this, consider the non-negative quantity $\<k|\Lambda(|i\>\<i|)|k\>\geq0$, where $\<k|i\>=0$ and $|i\>\<i|,|k\>\<k|\in\chi_C$. This implies $t-r\geq0$ from Eq.~\eqref{eq:comp_depolarizing_r_t}}, so $r$ is also sensitive to leakage errors.  Conceptually, leakage causes population to leave the computational space so it also causes phases within the computational space to be destroyed, and therefore computational errors. So it is more accurate to say that computational and leakage errors affect $r$ and leakage only affects $t$. Alternatively, computational errors only affect $\lambda$ (since we subtract out the leakage part from $r$) and leakage errors only affect $\tau$.

The total quality of a gate is quantified with the average fidelity. The average fidelity is defined as the average state fidelity over all pure states in the computational subspace
\begin{align} \label{eq:fidelities0}
    F &:=\int d \psi_C \bra{\psi_C} \Lambda \left(\ketbra{\psi_C}{\psi_C}\right) \ket{\psi_C}, \nonumber \\
    &=\int d \psi_C \bra{\psi_C} \bar{\Lambda}_{CC} \left(\ketbra{\psi_C}{\psi_C}\right) \ket{\psi_C}, \nonumber\\
    &= \frac{d_C-1}{d_C}r+\frac{1}{d_C}t, \nonumber\\
    &= 1 - \frac{d_C-1}{d_C}\lambda - \tau,
\end{align}
for $\ketbra{\psi_C}{\psi_C} \in \chi_C$. The second line uses the fact that the fidelities of $\Lambda$ and the twirled channel $\bar{\Lambda}$ are equal and the final line uses Eq.~\eqref{eq:comp_depolarizing_r_t}. A similar expression can be written for the process (or entanglement) fidelity, which is an alternative fidelity definition for gate quality, i.e.,
\begin{align} \label{eq:fidelities}
    f &= \frac{1}{d_C^2} \sum_{i=0}^{d_C^2-1} \textrm{Tr}[P_{C,i} \Lambda(P_{C,i})], \nonumber \\
    &= \frac{d_C^2-1}{d_C^2}r+\frac{1}{d_C^2}t, \nonumber \\
    &= 1 - \frac{d_C^2-1}{d_C^2}\lambda - \tau,
\end{align}
where $P_{C,i}=\tfrac{1}{\sqrt{d_C}}P_i \oplus 0 \in \chi_C$ is a computational Pauli operator and there is zero projection on the leakage subspace. By convention, we set $P_{C,0}:=\tfrac{1}{\sqrt{d_C}}\mathds{I}_C \oplus 0$ where $\mathds{I}_C$ is the identity in the computational operator space.
While it is best to measure both $r$ (or $\lambda$) and $t$ to estimate $F$ (or $f$), the estimate of $r$ serves as a reasonable bound on $F$ (or $f$),
\begin{align} \label{eq:fidelity_bound}
    r&\leq F \leq 1-\tfrac{d_C-1}{d_C}(1-r), \nonumber \\
     r &\leq f \leq 1 - \tfrac{d_C^2-1}{d^2_C}(1-r),
\end{align}
where the relation $r\leq t$ is used \footnotemark[2]. The fidelity $F$ and $f$ approach $r$ exponentially fast in the number of qubits ($d_C=2^N$).

\section{Revised randomized benchmarking with leakage} \label{sec:rb_methods}

In order to estimate fidelity with leakage from RB we need to measure parameters $\(r,t\)$ or $\(\lambda,\tau\)$ and properly account for the measurement operator's leakage projection. We propose three methods that accomplish these goals:

\textbf{Computational state preparation and measurement (Comp. SPAM):} In this method, we prepare an initial state in the computational space and apply a measurement operator that contains no population from leakage subspaces. Most systems aim to produce high fidelity computational states for quantum computing. The computational measurement may only be available in certain systems, since some measurement operators might contain projection onto leakage states (for example, atom loss with state-dependent resonance flouresence in dark states). Furthermore, we use rank-1 projectors for the initial state and the measurement, i.e., $\rho_{\textrm{in}}=\Pi_{\textrm{out}} = \ketbra{i}{i}_C\in\chi_C$, which is typical for most quantum computing systems. The survival probability under this setting is 
\begin{align}\label{eq:comp_SPAM_survival}
    p_{\textrm{comp}}(\ell)=\Tr\[\Pi_{\textrm{out}}\bar{\Lambda}^{\ell}(\rho_{\textrm{in}})\],
\end{align}
which will be used to obtain the error parameters $\(r,t\)$ or $\(\lambda,\tau\)$. 

\textbf{Average Over Measurement Basis (Avg. MB):} In this method, we include an additional unitary compiled into the final inversion gate to deterministically map each ideal output to a different measurement operator like in Ref.~\cite{Harper19}. We then average over the measurement outputs to evenly distribute all possible leakage projections from different measurement operators to avoid accidentally using a measurement operator that has a large or small leakage projection. For example, in a 1Q RB where the ideal noiseless output from a sequence is $|0\>\<0|$, the procedure is then to run the sequences twice: once with no final gate and targeted measurement operator $|0\>\<0|$ and once with an additional Pauli $X$ and targeted measurement $|1\>\<1|$. We then evaluate the survival probability averaging over both settings. By doing so, all leakage outcomes are included such that the survival decay provides direct information about $\(r,t\)$. This single qubit example is the procedure introduced in Ref.~\cite{Andrews19}, but we do not use heuristic decay ansatz as in Ref.~\cite{Andrews19} to extract leakage rate from it. Instead, we will show that this procedure gives derivable decay forms for extracting error parameters of \emph{multi-qubit} gates as was suggested, but not implemented, in Ref.~\cite{Weinstein23}.

More formally, a targeted computational measurement outcome $\Pi_k = |k\>\<k|_C\in\chi_C$ may actually have some projection onto the leakage subspaces $\Pi_k \rightarrow \ketbra{k}{k}_C + \Pi_{L,k}$ where $\Pi_{L,k} \in \chi_L$ is a (possibly unknown) combination of projectors onto the leakage subspace. By the definition of POVM, summing over all outcomes resolves the identity operator on the joint space, i.e., $\sum_k \Pi_k = \mathds{I} = \mathds{I}_C + \mathds{I}_L$ where $\sum_k \ketbra{k}{k}_C = \mathds{I}_C$ and $\sum_k \Pi_{L,k} = \mathds{I}_L$.

Suppose the initial state $\rho_{\textrm{in}}$ is a computational basis projector $\rho_{\textrm{in}}=|i\>\<i|_{\textrm{in}}\in\chi_C$. We label the additional unitary that is compiled into the final gate as $Q_k$, where $Q_k$ maps the operator $|k\>\<k|_C$ back to the ideal output $|i\>\<i|_{\textrm{in}}$ of the RB sequence, i.e., $Q_k^{\dagger}|k\>\<k|_CQ_k=|i\>\<i|_{\textrm{in}}$. Such $Q_k$ can be performed by 1Q Pauli $X$ and the identity gate on each qubit, i.e., $Q_k\in\{I,X\}^{\otimes N}$, which can be compiled into the final inversion gate. Operationally, for each $Q_k$ gate applied, we change the accepted ideal outcome to $|k\>\<k|_C$. The effective measurement operator becomes $Q_k^{\dagger}\Pi_k Q_k=|i\>\<i|_{\textrm{in}}+Q^{\dagger}_k\Pi_{L,k}Q_k$. We then evaluate the survival probability of averaging over all basis $k$, i.e.,
\begin{align}\label{eq:avg_basis_outcomes}
    &p_{\textrm{avg}}(\ell)=\frac{1}{d_C}\sum_{k}\Tr\[Q_k^{\dagger}\Pi_kQ_k\bar{\Lambda}^{\ell}(\rho_{\textrm{in}})\] \\
    &=\Tr\[\rho_{\textrm{in}}\bar{\Lambda}^{\ell}(\rho_{\textrm{in}})\]+\frac{1}{d_C}\sum_k\Tr\[Q^{\dagger}_k\Pi_{L,k}Q_k\bar{\Lambda}^{\ell}(\rho_{\textrm{in}})\]. \nonumber
\end{align}
In App.~\ref{app:leak_identity_approx}, we argue that the second term in the equation above is approximately equal to the leakage population after the RB sequence, i.e.,
\begin{align}
    \sum_k\Tr\[Q^{\dagger}_k\Pi_{L,k}Q_k\bar{\Lambda}^{\ell}(\rho_{\textrm{in}})\]\approx \Tr\[\mathds{I}_{L}\bar{\Lambda}^{\ell}(\rho_{\textrm{in}})\].
\end{align}
Then, the averaged survival probability becomes
\begin{align}\label{eq:avg_basis_survival}
    p_{\textrm{avg}}(\ell)=\Tr\[\rho_{\textrm{in}}\bar{\Lambda}^{\ell}(\rho_{\textrm{in}})\]+\frac{\Tr\[\mathds{I}_{L}\bar{\Lambda}^{\ell}(\rho_{\textrm{in}})\]}{d_C},
\end{align}
which we will use to derive explicit functional forms in terms of error parameters $\(r,t\)$.

In addition, we find that a different setting is required to extract the information about the leakage rate $\tau$ (or $t$). In this setting, we require a particular measurement operator that contains no projections onto the leakage subspaces. Let $\Pi_{\textrm{out}}=|i\>\<i|_{C}\in\chi_C$ be such a measurement operator. We perform the additional unitary $Q_k$ at the end of the RB sequence and measure $\Pi_{\textrm{out}}$ for every setting. The effective measurement is $Q^{\dagger}_k\Pi_{\textrm{out}}Q_k$, and summing over all $k$ gives the survival probability
\begin{align}\label{eq:comp_identity_survival}
    p_{\mathds{I}_C}(\ell)=\sum_k\Tr\[Q^{\dagger}_k\Pi_{\textrm{out}}Q_k\bar{\Lambda}^{\ell}(\rho_{\textrm{in}})\]=\Tr\[\mathds{I}_C\bar{\Lambda}^{\ell}(\rho_{\textrm{in}})\],
\end{align}
where $\mathds{I}_C$ is the identity operator in the computational subspace $\chi_C$ and we call this the computational identity measurement setting.

\begin{figure}
    \centering
    \includegraphics[width=\columnwidth]{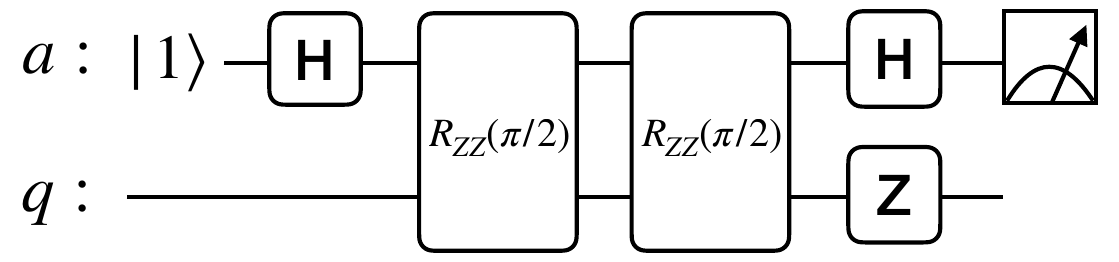}
    \caption{The leakage gadget uses an ancilla qubit ``$a$”  to detect if a target qubit ``$q$” has
leaked. The ancilla is initially prepared in $|1\>$. If ``$q$” has leaked, the 2Q gates have no effect, and ``$a$” is measured as $|1\>$. If
instead ``$a$” has leaked, then ``$a$” will also be measured as $|1\>$ since the
leakage is measured in a bright state. If neither ``$q$” nor ``$a$”
has leaked, then ``$a$” is measured as $|0\>$.}
    \label{fig:gadget}
\end{figure}

\textbf{Leakage post-selection (LPS):} In this method, we include an additional measurement of the leakage population in order to differentiate leakage from computational errors. Leakage population can be measured in at least two ways: (1) separately addressing the states either with shelving or detuning such that the final measurement is described by separate projection onto each computational and leakage states, or (2) with a leakage gadget (e.g. the one shown in Fig.~\ref{fig:gadget}) that uses ancilla qubits~\cite{Stricker2020,Moses23}.

We will show that in many cases, by post-selecting the shots from the dataset, we can isolate $\lambda = t - r$, and by fitting the data retention rate (the fraction of shots we keep), we can isolate $t$. Explicitly, the data retention rate is collecting all outcomes that contain no leakage population, i.e.,
\begin{align}\label{eq:retention_survival}
    p_{\textrm{retention}}(\ell)=\Tr\[\mathds{I}_C\bar{\Lambda}^{\ell}(\rho_{\textrm{in}})\],
\end{align}
which serves as the same purpose of the computational identity measurement in Eq.~\eqref{eq:comp_identity_survival} and relates to $t$ or $\tau$ directly for the cases we consider in the following section. In addition, we can use the post-selected outcomes to achieve a computational measurement $\Pi_{\textrm{out}}\in\chi_C$, similar to the same effect of the computational SPAM setting where $\Pi_{\textrm{out}}=\rho_{\textrm{in}}\in\chi_C$. The post-selected survival probability becomes
\begin{align}\label{eq:post_selected_survival}
    p_{\textrm{post}}(\ell)=\frac{\Tr\[\Pi_{\textrm{out}}\bar{\Lambda}^{\ell}(\rho_{\textrm{in}})\]}{p_{\textrm{retention}}(\ell)}.
\end{align}
Combining the decay information from $p_{\textrm{retention}}(\ell)$ and $p_{\textrm{post}}(\ell)$ allows us to obtain both $\(r,t\)$ or $\(\lambda,\tau\)$ and hence the fidelity $F$.

Post-selection is used in both Refs.~\cite{Radnaev24,Muniz24} to account for atom loss and leakage. However, they use heuristic modeling of the error process similar to simple probabilistic error counting and its applicability is unknown for general cases. In the next section, we show how to use post-selection with the carefully chosen sequence lengths for each error regime to properly measure leakage and the fidelity. 

A prospective benchmarker will need to asses their systems capability to determine which method is best suited. For example, not all systems have the ability to implement a leakage gadget so LPS may not be viable but if a computational only measurement operator exists then Avg. MB and Comp. SPAM are valid. As we see in numerical simulations, some methods also perform better in certain error regimes depending on the relative magnitude of computational and leakage errors.

\begin{table*}[]
\begin{tabular}{|l|c|c|c|}
\hline
& Comp. SPAM & Avg. MB & LPS \\ \hline \hline
\begin{tabular}[c]{@{}l@{}}Short lengths\\ Sec.~\ref{sec:small} \\ $\lambda \ell \ll 1, \tau \ell \ll 1$ \end{tabular}   & $p_{\textrm{comp}}(\ell)=1-\ell\left(\frac{d_C-1}{d_C}\lambda+\tau \right)$ & \begin{tabular}[c]{@{}c@{}} $p_{\textrm{avg}}(\ell)=1-\ell\tfrac{d_C-1}{d_C}(\lambda+\tau)$ \\ $p_{\mathds{I}_C}(\ell)= 1-\ell \tau$ \end{tabular} & \begin{tabular}[c]{@{}c@{}} $p_{\textrm{post}}(\ell)=1-\ell\tfrac{d_C-1}{d_C}\lambda$ \\ $p_{\textrm{retention}}(\ell)= 1-\ell \tau$ \end{tabular} \\
\hline
\begin{tabular}[c]{@{}l@{}}Comp. dominant\\ Sec.~\ref{sec:comp_dominant} \\ $\tau \ell \ll 1$\end{tabular}  & $p_{\textrm{comp}}(\ell)=\frac{d_C-1}{d_C}(1 - \lambda - \ell \tau)(1 - \lambda)^{\ell-1} + \frac{1- \ell \tau}{d_C}$ & \begin{tabular}[c]{@{}c@{}} $p_{\textrm{avg}}(\ell)=\frac{d_C-1}{d_C}( 1- \lambda - \tau)^{\ell} + \frac{1}{d_C}$ \\ $p_{\mathds{I}_C}(\ell)= 1-\ell \tau$ \end{tabular} & \begin{tabular}[c]{@{}c@{}} $p_{\textrm{post}}(\ell)=\frac{d_C-1}{d_C}(1-\lambda)^{\ell} + \frac{1}{d_C}$ \\ $p_{\textrm{retention}}(\ell)= 1-\ell\tau$ \end{tabular} \\
\hline
\begin{tabular}[c]{@{}l@{}}No seepage\\ Sec.~\ref{sec:no_seepage} \\ $\bar{\Lambda}_{CL} = 0$ \end{tabular}  & $p_{\textrm{comp}}(\ell)=\frac{d_C-1}{d_C}r^{\ell} + \frac{1}{d_C}t^{\ell}$ & \begin{tabular}[c]{@{}c@{}} $p_{\textrm{avg}}(\ell)=\frac{d_C-1}{d_C}r^{\ell} + \frac{1}{d_C}$ \\ $p_{\mathds{I}_C}(\ell)= t^{\ell}$ \end{tabular} & \begin{tabular}[c]{@{}c@{}} $p_{\textrm{post}}(\ell)=\frac{d_C-1}{d_C}( \tfrac{r}{t})^{\ell} + \frac{1}{d_C}$ \\ $p_{\textrm{retention}}(\ell)= t^{\ell}$ \end{tabular} \\
\hline
\begin{tabular}[c]{@{}l@{}}Population transfer\\ Sec.~\ref{sec:pop_transfer} \end{tabular}  & - & $p_{\textrm{avg}}(\ell)=\frac{d_C-1}{d_C}r^{\ell} + \frac{1}{d_C}$  & - \\
\hline
\end{tabular}
\caption{Table for all error regimes described in subsequent sections and corresponding RB schemes and survival probabilities. \label{tab:error_methods}}
\end{table*}

\section{Survival probability derivations and simulations} \label{sec:survival_probs}
In this section, we derive survival probabilities for three different error regimes with the three different RB methods from Sec.~\ref{sec:rb_methods}. Here, we do not consider the effects of SPAM to ease the notation and postpone its discussion in App.~\ref{app:SPAM_surv}. We also show how standard RB breaks down when the selected measurement bases contains projections onto the leakage states in Sec.~\ref{sec:when_leakage_ignored}.

The three error regimes we consider are: (1) short sequences such that more than one computational or leakage errors are unlikely, (2) computational dominant errors such that more than one leakage error is unlikely, (3) no seepage, and (4) population transfer. The first two relate to the relative size of the errors and require special sequence length selection. The second two relax the relative size restrictions but require more assumptions about the structure of the error. A perspective benchmarker with no knowledge of the error could potentially start running short sequences to get an early estimate and then move to one of the other three regimes based on that measurement or other knowledge of the system. 

In the numerical simulations, we simulate a set of 2Q RB experiments on a 2Q leakage error process with magnitude $\tau_s$, combined with a computational error process with magnitude $\lambda_s$ to demonstrate the effectiveness of each method. We select sequence lengths that respect each error regime based on the values $\lambda_s$ and $\tau_s$ given by the rules in Table~\ref{tab:RB_lengths}. For each error regime, we probe a range $\tau_s$ and $\lambda_s$ and create a heat-map plot of the relative difference $|x-x_s|/x_s$ for injected error process parameter $x_s$ and measured error parameter $x$ for parameters $1-F$, $1-r$~\footnote[3]{It might be more useful in some contexts to estimate $\lambda$ but several methods directly estimate $r$ and then extracting $\lambda$ has higher uncertainty due to the uncertainty in $r$ and $t$ adding but the estimate coming from the difference $\lambda = t-r$. With $r$ or $\lambda$ we can still estimate the fidelity by scaling the factors and adding with $t$}, and $1-t$. Each heat-map plot has an absolute scale of 0 to 1 in all error regimes for clarity, although the relative difference can be greater than one in some cases and we note that in the respective subsections. Each error regime and method may require different circuit runtimes and resources due to the different sequence length requirements. We do not attempt to compare these tradeoffs since resource runtime is system specific. More details about the example process and sequence lengths are given in App.~\ref{app:numerics}. 

\subsection{Short sequences} \label{sec:small}
The first error regime we consider is running short RB sequences such that the probability of having more than one error per RB sequence is negligible. First, we decompose the error process into the identity part and the error part, i.e.,
\begin{align} \label{eq:1o_process_expansion}
    \bar{\Lambda}&=\mathcal{I}+\begin{pmatrix}
        -(\lambda+\tau)\mathcal{I}_C + \tfrac{\lambda}{d_C}\Tr\[\mathcal{I}_C(\cdot)\]\mathds{I}_C & \bar{\Lambda}_{CL} \\
        \bar{\Lambda}_{LC} & \mathcal{E}_{LL}
    \end{pmatrix} \nonumber\\
    &:=\mathcal{I}+\mathcal{E},
\end{align}
where the ideal operation is $\mathcal{I}$ and $\mathcal{E}$ corresponds to the added error per gate such that $||\mathcal{E}||=0$ when there is no error. Errors are small for sequence lengths such that $\ell||\mathcal{E}||\ll 1$. The error process after applying $\ell$ gates in the RB sequence is
\begin{align} \label{eq:1o_process}
    \bar{\Lambda}^{\ell}=\(\mathcal{I}+\mathcal{E}\)^{\ell}\approx \mathcal{I} +\ell \mathcal{E}.
\end{align}

\subsubsection{RB when leakage is ignored}\label{sec:when_leakage_ignored}
Before applying the methods from Sec.~\ref{sec:rb_methods}, let us first consider what happens when leakage is ignored, i.e., assuming $\tau=0$. Suppose we perform RB in the short sequence regime as described in Eq.~\eqref{eq:1o_process}, the decay of the average survival probability is
 \begin{align} \label{eq:standard_decay}
    p(\ell) &= \Tr\left[\Pi_{\textrm{out}} \bar{\Lambda}^{\ell}(\rho_{\textrm{in}}) \right]\nonumber \\
     &\approx 1 + \ell \Tr\left[\Pi_{\textrm{out}} \mathcal{E}(\rho_{\textrm{in}}) \right] = 1 - \ell  \frac{d_C-1}{d_C} \lambda,
\end{align}
where the slope is treated as the infidelity $1-F=\tfrac{d_C-1}{d_C}\lambda$. 

As shown in Fig.~\ref{fig:leakage_RB_comparisons}, even if leakage is non-zero but small, the effect on the estimated fidelity may be substantial depending on the final measurement operator selection. To see this, suppose that the final measurement operator contains some leakage outcomes, i.e., $\Pi_{\textrm{out}} = \rho_{\textrm{in}}+\Pi_L$ where $\rho_{\textrm{in}}\in\chi_C$ and $\Pi_L \in \chi_L$. In the short sequence limit, the survival probability including leakage becomes
\begin{align}
    p(\ell)&= \Tr\left[\Pi_{\textrm{out}} \bar{\Lambda}^{\ell}(\rho_{\textrm{in}}) \right]\nonumber\\
    &\approx 1+ \ell \Tr\left[\Pi_{\textrm{out}} \mathcal{E}(\rho_{\textrm{in}}) \right], \nonumber \\
     &= 1- \ell \left( \frac{d_C - 1}{d_C}\lambda + \tau\right) +  \ell\Tr\left[\Pi_L \bar{\Lambda}_{LC}(\rho_{\textrm{in}} )\right].
\end{align}
In the extreme case where the measurement $\Pi_{\textrm{out}}$ contains all the leakage outcomes $\Pi_L = \mathds{I}_L$, we have $\Tr\left[\mathds{I}_L \bar{\Lambda}_{LC}(\rho_{\textrm{in}} )\right]= \tau$. The survival probability as a function of $\ell$ becomes $p(\ell) = 1 - \tfrac{d_C-1}{d_C} \ell \lambda$, where the slope only measures the computational error $\lambda$ and reporting it as the infidelity then leads to an inaccurate estimate. As an example, consider the case where the total error is dominated by leakage, e.g., $\lambda\approx0$ and $\tau=0.1$. then the estimated infidelity in this method would be zero while the actual infidelity is $1-F=\tau=0.1$, a rather large difference!

\subsubsection{Computational SPAM}
To properly measure the fidelity, we first use the Comp. SPAM method provided in Sec.~\ref{sec:rb_methods} and the short sequence approximation from  Eq.~\eqref{eq:1o_process} to derive survival probability for a rank-1 initial state and the measurement operator $\rho_{\textrm{in}}=\Pi_{\textrm{out}}\in \chi_C$, i.e.,
\begin{align} \label{eq:linear-survival}
    p_{\textrm{comp}}(\ell)=1-\ell\left(\frac{d_C-1}{d_C}\lambda+\tau \right). 
\end{align}
The slope of the decay gives the infidelity $1-F$, as defined in Eq.~\eqref{eq:fidelities0}. This differs from the previous subsection since the measurement only contains computational projection by design. 

\subsubsection{Average over measurement basis}\label{sec:short_seq_avg_meas}
One can also use the Avg. MB method in Sec.~\ref{sec:rb_methods} to derive an averaged survival decay $p_{\textrm{avg}}(\ell)$ from Eq.~\eqref{eq:avg_basis_survival} with the linear approximation of the error process in Eq.~\eqref{eq:1o_process}. Explicitly, we have
\begin{align}
    p_{\textrm{avg}}(\ell)=1-\ell\(\tfrac{d_C-1}{d_C}\lambda+\tau\)+\tfrac{\ell}{d_C}\Tr\[\mathds{I}_L\bar{\Lambda}_{LC}(\rho_{\textrm{in}})\].
\end{align}
From the definition in Eq.~\eqref{eq:lambda}, we have
\begin{equation}
    \Tr\[\mathds{I}_L\bar{\Lambda}_{LC}(\rho_{\textrm{in}})\] 
    =\Tr\[\mathds{I}_L\Lambda_{LC}\(\tfrac{\mathds{I}_C}{d_C}\)\]=\tau,
\end{equation}
where the final equality comes from the definition of the leakage rate $\tau$ in Eq.~\eqref{eq:t} and Ref.~\cite{Wood18} and we assume $\Lambda$ is trace preserving over the full Hilbert space. Therefore, the average survival probability over all measurement operators is
\begin{align}
    p_{\textrm{avg}}(\ell)=1-\ell\tfrac{d_C-1}{d_C}(1-r).\label{eq:avg_over_readout_1o}
\end{align}
To measure the leakage rate $\tau$, we can perform the computational identity measurement introduced in Avg. MB of Sec.~\ref{sec:rb_methods} to all states and sum over the outputs. Specifically, combining Eq.~\eqref{eq:comp_identity_survival} and Eq.~\eqref{eq:1o_process} gives
\begin{align}
    &p_{\mathds{I}_C}(\ell)= 1+ \ell \Tr\[\mathds{I}_C\mathcal{E}(\rho_{\textrm{in}})\]= 1-\ell \tau.
\end{align}
The slope of the decay gives the leakage rate $\tau$. Combining with the slope from Eq.~\eqref{eq:avg_over_readout_1o}, gives the fidelity $F$ from Eq.~\eqref{eq:fidelities0}.

\subsubsection{Leakage post-selection}
For the LPS method in the short sequence regime, the data retention rate becomes
\begin{equation}
    p_{\textrm{retention}}(\ell)=\Tr\[\mathds{I}_C\(\mathcal{I}+\ell\mathcal{E}\)(\rho_{\textrm{in}})\]= 1 - \ell \tau. \label{eq:tau_short_LPS}
\end{equation}
where the slope of the decay gives the leakage rate $\tau$. If we use Comp. SPAM method from Eq.~\eqref{eq:linear-survival} by post-selecting on outcomes that are reported as no leakage, then
\begin{equation}
    p_{\textrm{post}}(\ell) = \frac{p_{\textrm{comp}}(\ell)}{p_{\textrm{retention}}(\ell)} = 1 - \ell\frac{d_C-1}{d_C} \lambda, 
\end{equation}
where the slope gives $\lambda$. Combining $\tau$ obtained from $p_{\textrm{retention}}(\ell)$ gives the fidelity $F$ from Eq.~\eqref{eq:fidelities0}.

\subsubsection{Simulations}

\begin{figure*} 
\centering
\includegraphics[width=\textwidth]{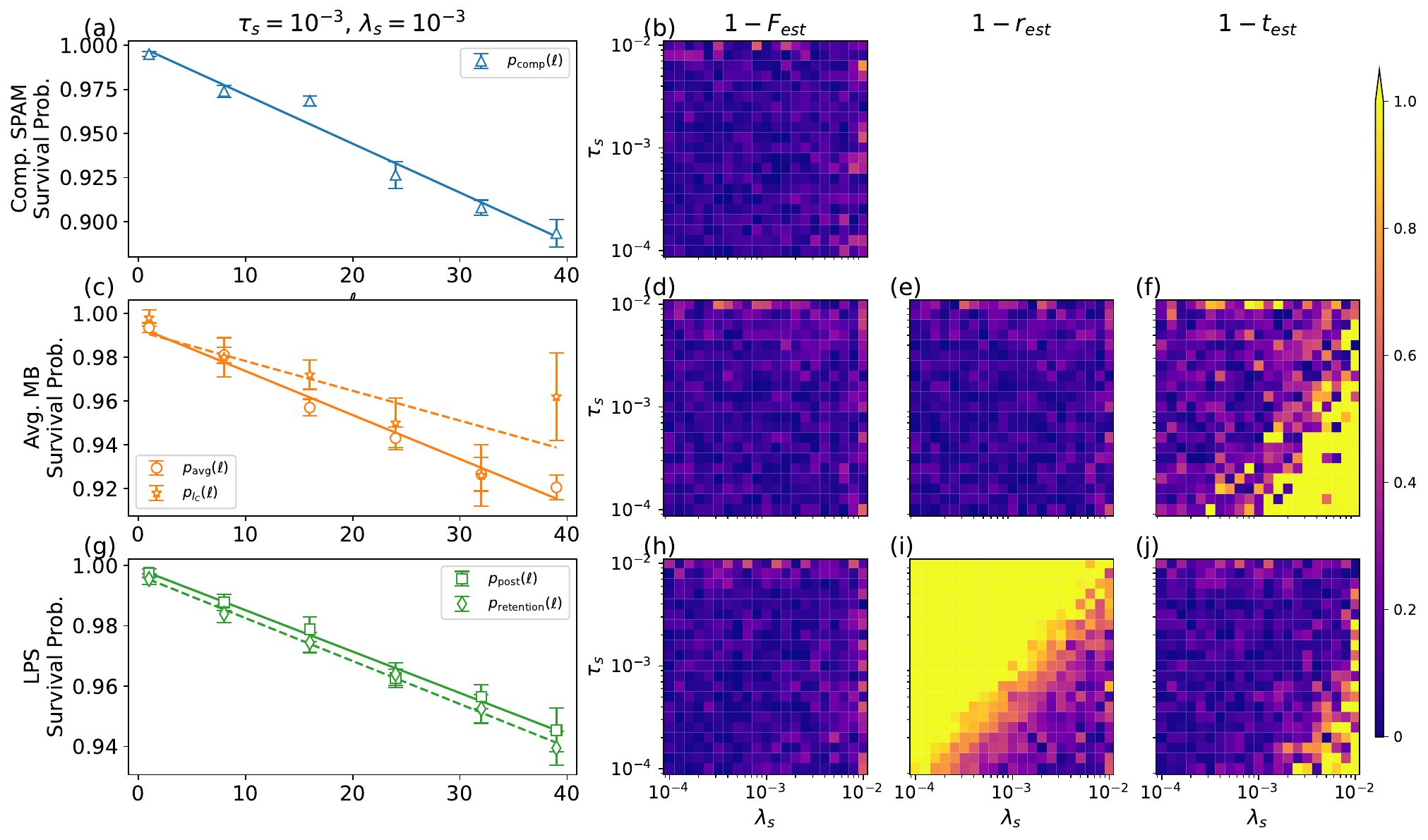}
\caption{Heat-map plots of the relative difference between the estimated values from each method compared to the values of the input error model using methods tailored to short sequences (e.g. $\tau \ell \ll 1, \lambda \ell \ll 1$). The x-axis of each subplot shows the injected value of $\lambda_s$ (the magnitude of the computational error) and the y-axis shows the injected value of $\tau_s$ (the magnitude of the leakage error). (a) Example decay curve for $\lambda_s=\tau_s=10^{-3}$ for the Comp. SPAM method, (b) Infidelity $1-F$ for the Comp. SPAM method, (c) Example decay curve for $\lambda_s=\tau_s=10^{-3}$ for the Avg. MB method (d) infidelity $1-F$ for the Avg. MB method, (e) $1-r$ for the Avg. MB method, (f) $\tau$ for the Avg. MB method, (g) Example decay curve for $\lambda_s=\tau_s=10^{-3}$ for the LPS method (h) infidelity $1-F$ for the LPS method, (i) $1-r$ for the LPS method, and (j) $\tau$ for the LPS method.\label{fig:small_errors} }
\end{figure*}

An example 2Q RB fit in this error regime is shown Figs.~\ref{fig:small_errors}a, c, and g for a single error process with $\tau_s=\lambda_s=10^{-3}$. Results of a larger simulation for error process with a range of $\tau_s$ and $\lambda_s$ values are shown in the other parts of Fig.~\ref{fig:small_errors}. We see that the maximum disagreement in $1-F$ between the estimated and the true infidelity of 0.75 for Comp. SPAM, 0.64 for Avg. MB, and 0.56 for LPS. These occur when there was a relatively large difference between $\lambda_s$ and $\tau_s$ (either is $\approx 10^{-2}$). In this regime, it is difficult to choose small enough sequence lengths that enforce the short sequence approximation causing larger contributions from shot noise. This occurs near the far right and top sides in all $1-F$ figures (Fig.~\ref{fig:small_errors}b, d and h). This also affects the estimates of $\tau$ from Avg. MB and LPS (far right of Fig.~\ref{fig:small_errors}f and j). Avg. MB is especially sensitive in this regime due to finite sampling effects that limit the precision of $\tau$ estimate since the computational population may sum to above when summing over  final unitary $Q_k$. The LPS method also produces poor estimates of $r$ when $\tau_s > \lambda_s$ (Fig.~\ref{fig:small_errors}i). In this regime, leakage dominates and much of the data is thrown out in post-selection so that the signal from $\lambda_s$ is difficult to resolve. Except for those specific situations, all three methods provide reliable estimates for most cases.

The advantage of this method is that it allows us to directly extract the fidelity and requires no further assumption about the details of the error process or the relative sizes between each error component. The disadvantage is that it requires using short sequences. There are reasons why it might not be ideal to only use short sequences in RB. For one, RB works by running long sequences to amplify errors making them easier to resolve from SPAM errors and shorter sequences have less resolution. Our numerical data does observe it becomes less shot efficient in this regime, where shot noise becomes a major source of error in some cases.

\subsection{Dominating computational error}\label{sec:comp_dominant}
The next error regime we consider is when the total error is dominated by computational errors and leakage errors are relatively small, i.e., $\lambda > \tau$. This is often the case in atomic gates where leakage and seepage are due to spontaneous emission that may be a small fraction of the total error budget. To treat this, we perform a perturbative expansion with respect to leakage and seepage while keeping the computational error exact. Conceptually, this allows us to use longer sequences $\ell$ where the probability of more than one leakage error is small but the probability of more than one computational error is unconstrained. Importantly, even when $\lambda> \tau$ leakage can still be a significant contribution to fidelity based on Sec.~\ref{sec:when_leakage_ignored}. 

To derive the survival probability in this regime, first, expand the error process to separate computational errors from leakage errors as
\begin{align} \label{eq:comp_dominant_expansion}
    \bar{\Lambda}&=\underbrace{\begin{pmatrix}
        (1-\lambda)\mathcal{I}_{C}+\tfrac{\lambda}{d_C}\Tr\[\mathcal{I}_C(\cdot)\]\mathds{I}_C &0 \\
        0 & \mathcal{I}_L
    \end{pmatrix}}_{:=\Lambda_{\textrm{comp}}}+\underbrace{\begin{pmatrix}
        -\tau\mathcal{I}_C &\bar{\Lambda}_{CL} \\
        \bar{\Lambda}_{LC} & \mathcal{E}_{LL}
    \end{pmatrix}}_{:=\mathcal{E}'},
\end{align}
where $\Lambda_{\textrm{comp}}$ is the computation error process and $\mathcal{E}'$ describes the error map due to leakage and seepage. After applying this process $\ell$ times, we only keep the terms involving up to one $\mathcal{E}'$ (i.e., under the assumption $\ell ||\mathcal{E}'||\ll 1$), namely
\begin{align} \label{eq:dominating_comp_expansion}
    &\bar{\Lambda}^{\ell}=\(\Lambda_{\textrm{comp}}+\mathcal{E}'\)^{\ell} \approx \Lambda_{\textrm{comp}}^{\ell}+ \sum_{k=1}^{\ell}\Lambda_{\textrm{comp}}^{\ell-k}\mathcal{E}'\Lambda_{\textrm{comp}}^{k-1}.
\end{align}

\begin{figure*} 
\centering
\includegraphics[width=\textwidth]{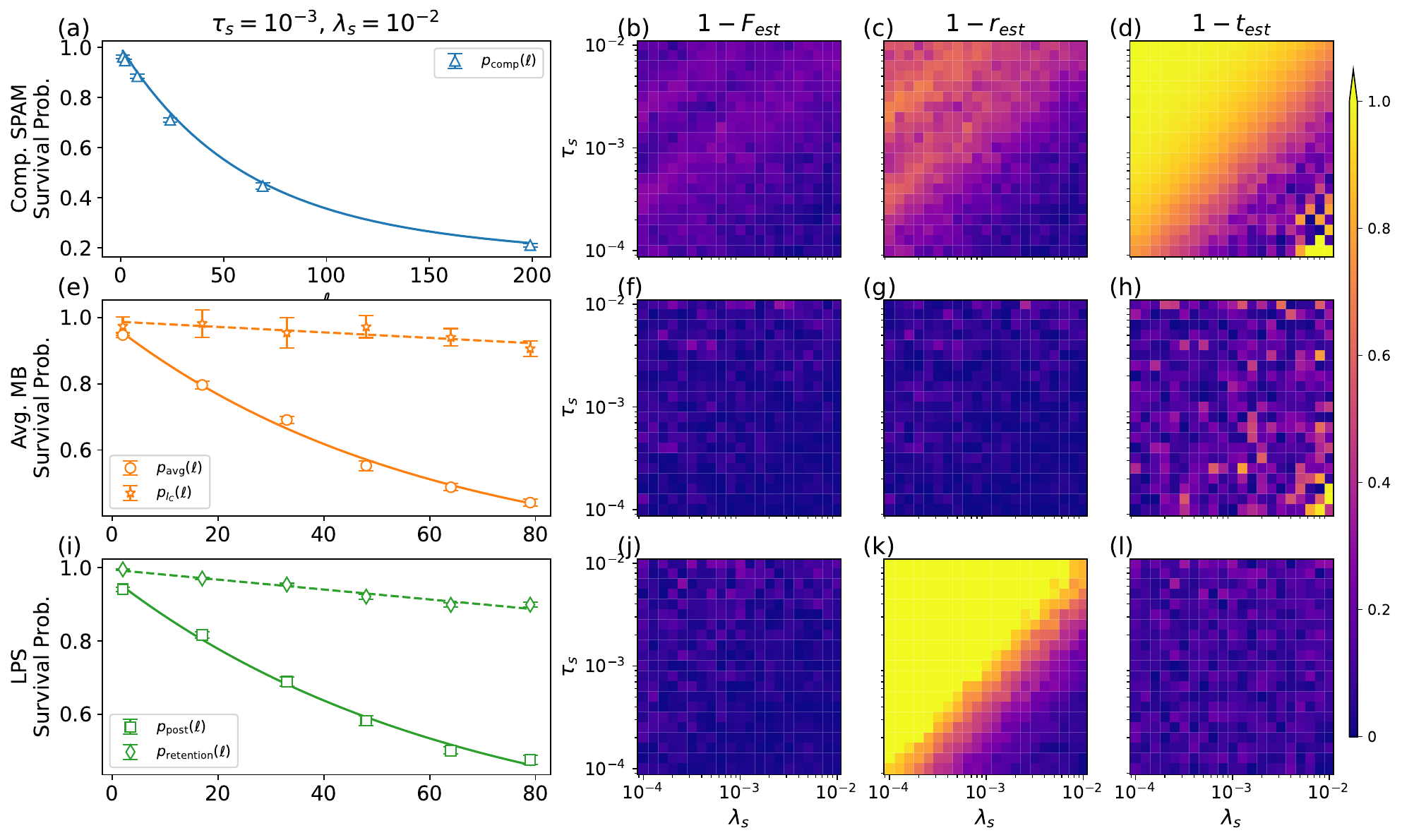}
\caption{Heat-map plots of the relative difference between the estimated values from each method compared to the values of the input error model using methods tailored to dominant computational errors (e.g. $\tau \ell \ll 1$ and $\tau < \lambda$). The x-axis of each subplot shows the injected value of $\lambda_s$ (the magnitude of the computational error) and the y-axis shows the injected value of $\tau_s$ (the magnitude of the leakage error). (a) Example decay curve for $\lambda_s=10^{-2},\tau_s=10^{-3}$ for the Comp. SPAM method, (b) Infidelity $1-F$ for the Comp. SPAM method, (b) $1-r$ for the Comp. SPAM method, (b) $\tau$ for the Comp. SPAM method, (c) Example decay curve for $\lambda_s=10^{-2},\tau_s=10^{-3}$ for the Avg. MB method (d) infidelity $1-F$ for the Avg. MB method, (e) $1-r$ for the Avg. MB method, (f) $\tau$ for the Avg. MB method, (g) Example decay curve for $\lambda_s=10^{-2},\tau_s=10^{-3}$ for the LPS method (h) infidelity $1-F$ for the LPS method, (i) $1-r$ for the LPS method, and (j) $\tau$ for the LPS method.\label{fig:comp_dominant} }
\end{figure*}

\subsubsection{Computational SPAM}
Here we use the Comp. SPAM method introduced in Sec.~\ref{sec:rb_methods}, where we prepare and measure in the computational space $\rho_{\textrm{in}}=\Pi_{\textrm{out}}=|k\>\<k|\in\chi_C$. The survival probability becomes 
\begin{align} \label{eq:comp_dominant_survival_prob}
    p_{\textrm{comp}}(\ell)= \frac{d_C-1}{d_C}(1-\lambda-\ell \tau)(1-\lambda)^{\ell-1} +\frac{1-\ell \tau}{d_C}.
\end{align}
See App. \ref{app:comp_large_eq} for a full derivation. By fitting measured decay rates to Eq.~\eqref{eq_app:comp_identiy_leak_surv}, we can individually extract $\lambda$ and $\tau$, and deduce the average fidelity $F$ from Eq.~\eqref{eq:fidelities0}. 

Comp. SPAM in this error regime does require sufficiently long sequence lengths $\ell$ to be able to reliably fit $\tau$ and $\lambda$ from the above functional form. If $\ell$ is too short comparing to $\lambda$, then the survival reduces to the short sequence case discussed before where the slope can not differentiate $\tau$ and $\lambda$.  Based on the numerical tests, we find that choosing $\ell\lambda>1$ and $\ell\tau \ll1$ gives the best results.

\subsubsection{Average over measurement basis}

Alternatively, we can use the Avg. MB method described in Sec.~\ref{sec:rb_methods} where averaging over measurement basis gives
\begin{align}
    p_{\textrm{avg}}(\ell)= \frac{d_C-1}{d_C}r^{\ell}+\frac{1}{d_C}.
\end{align}
See App. \ref{app:comp_large_eq} for a full derivation. To obtain the leakage rate $\tau$, we use the computational identity measurement from Eq.~\eqref{eq:comp_identity_survival}, i.e.,
\begin{align}\label{eq:comp_dom_identity_meas}
    p_{\mathds{I}_C}(\ell)=\Tr\[\mathds{I}_C\bar{\Lambda}^{\ell}(\rho_{\textrm{in}})\] = 1-\ell \tau,
\end{align}
where the slope gives $\tau$. Detailed derivation is in App.~\ref{app:comp_large_eq}. Therefore, $\tau$ and $\lambda$ can be separately obtained to deduce the fidelity $F$ from Eq.~\eqref{eq:fidelities0}.

\subsubsection{Leakage post-selection}
Using the LPS method in Sec.~\ref{sec:rb_methods}, the post-selected survival probability is
\begin{align} 
    p_{\textrm{post}}(\ell)=\frac{d_C-1}{d_C}(1-\lambda)^{\ell}+\frac{1}{d_C},
\end{align}
and the data retention rate is 
\begin{align} \label{eq:comp_dominant_retention_survival_prob}
    p_{\textrm{retention}}(\ell)= 1- \ell \tau.
\end{align}
See App.~\ref{app:comp_large_eq} for detail. Therefore, one can obtain $\lambda$ and $\tau$ separately from $p_{\textrm{post}}(\ell)$ and $p_{\textrm{retention}}(\ell)$, and deduce the fidelity $F$ from Eq.~\eqref{eq:fidelities0}.

\subsubsection{Simulations}

Results of a larger simulation for error process with a range of $\tau_s$ and $\lambda_s$ values are shown in Fig.~\ref{fig:comp_dominant}. We see that the maximum relative difference between the estimated and the true infidelity  is 0.31 for the Comp. SPAM method, 0.31 for Avg. MB, and 0.27 for the LPS method. All occur when $\lambda_s \ll \tau_s$ (upper left half of each plot), which violates the computational dominant assumption $\lambda_s\gg\tau_s$. In each leakage plot, Fig.~\ref{fig:comp_dominant}d, h, and l there are values that have relative difference $>1$ that have been truncated for clarity. For the Comp. SPAM method, there are leakage estimates with $\lambda_s \gg \tau_s$ that have a large percent difference. We attribute this to the difficulty of fitting small values of $\tau$ with the number of shots used in the simulation. In other tests, we see that adding more shots or sequences alleviates this problem. For the Comp. SPAM fit we do not use SPAM fit parameters and despite having SPAM errors in the simulation we still see good agreement with infidelity estimates. 

This error regime alleviates some of the concerns from the previously considered short sequence regime but has more complicated survival probability functions making fitting more challenging. The Avg. MB and LPS methods have simpler fit functions but the Avg. MB has more assumptions and the LPS method has the additional problem that more data is thrown away with longer sequences making the measured survival probabilities noisier.

\subsection{No seepage} \label{sec:no_seepage}
The previous error regimes focused on the situation where the relative scale between the computation and leakage error is \textit{a priori} known and is used to select sequence lengths that allow for derivations of survival probabilities. Here, we consider a situation where there is no restriction on the sizes of the computation and leakage error but seepage is negligible. This can be satisfied in many atomic systems either due to atom loss (i.e., the atom with the encoded qubit is ejected from its trapping potential and can no longer interact) or from leaking population to other atomic sublevels that do not couple well to the qubit subspace. 

Without seepage $\Lambda_{CL}=0$, the error process becomes
\begin{align}
    \bar{\Lambda} =  
    \begin{pmatrix}
        \bar{\Lambda}_{CC} & 0\\
        \bar{\Lambda}_{LC} & \bar{\Lambda}_{LL}
    \end{pmatrix}.
\end{align}
Due to the special form of the process, the net process of a length $\ell$ sequence of Clifford gates becomes
\begin{equation} \label{eq:no_seepage_expansion}
    \bar{\Lambda}^{\ell} =  
    \begin{pmatrix}
        \bar{\Lambda}_{CC}^{\ell} & 0\\
        Y & \bar{\Lambda}_{LL}^{\ell}
    \end{pmatrix},
\end{equation}
where $Y$ is some complicated operator dependent on $\ell$, $\bar{\Lambda}_{CC}$, $\bar{\Lambda}_{LC}$, and $\bar{\Lambda}_{LL}$. Note that the upper-left block of the error process $\bar{\Lambda}^{\ell}_{CC}$ still allows for an analytic expression which we can derive survival probabilities with.

This regime was considered in Ref.~\cite{Wallman15} where a specific procedure was produced to measure $\tau$ only. This procedure did not use the final inversion gate like in standard RB and therefore could not also estimate $\lambda$ (or $r$), and therefore fidelity.

\subsubsection{Computational SPAM}\label{sec:2exp}
Using the Comp. SPAM method in Sec.~\ref{sec:rb_methods}, the survival probability for a computational initial state and measurement  $\rho_{\textrm{in}}=\Pi_{\textrm{out}}=|\psi\>\<\psi| \in \chi_C$ is
\begin{equation} \label{eq:general_decay}
    p_{\textrm{comp}}(\ell) = \Tr\[ \Pi_{\textrm{out}} \bar{\Lambda}^{\ell}\(\rho_{\textrm{in}}\)\]= \Tr\[ \Pi_{\textrm{out}} \bar{\Lambda}_{CC}^{\ell}\(\rho_{\textrm{in}}\) \]. \nonumber
\end{equation}
Applying $\ell$ times the error process $\bar{\Lambda}_{CC}$ gives
\begin{equation}\label{eq:lambda_CC_to_the_ell}
    \bar{\Lambda}_{CC}^{\ell}=r^{\ell}\mathcal{I}_C+\(t^{\ell}-r^{\ell}\)\Tr\[\mathcal{I}_C(\cdot)\]\mathds{I}_C/d_C
\end{equation}
as shown in App.~\ref{app:proof_for_lambda_to_l}. The survival probability becomes
\begin{align} \label{eq:2exp_fit}
p_{\textrm{comp}}(\ell)=\frac{d_C-1}{d_C}r^{\ell}+\frac{1}{d_C}t^{\ell}.
\end{align}
One can obtain $r$ and $t$ separately by fitting the decay curve with this two-exponential function and recover the average fidelity with $F$ from Eq.~\eqref{eq:fidelities0}. In order to enhance the stability of the fit, one may replace $r\to 1-\tau-\lambda$ and $t\to 1-\tau$ so that the constraint  $r\leq t$ is always imposed.

\begin{figure*} 
\centering
\includegraphics[width=\textwidth]{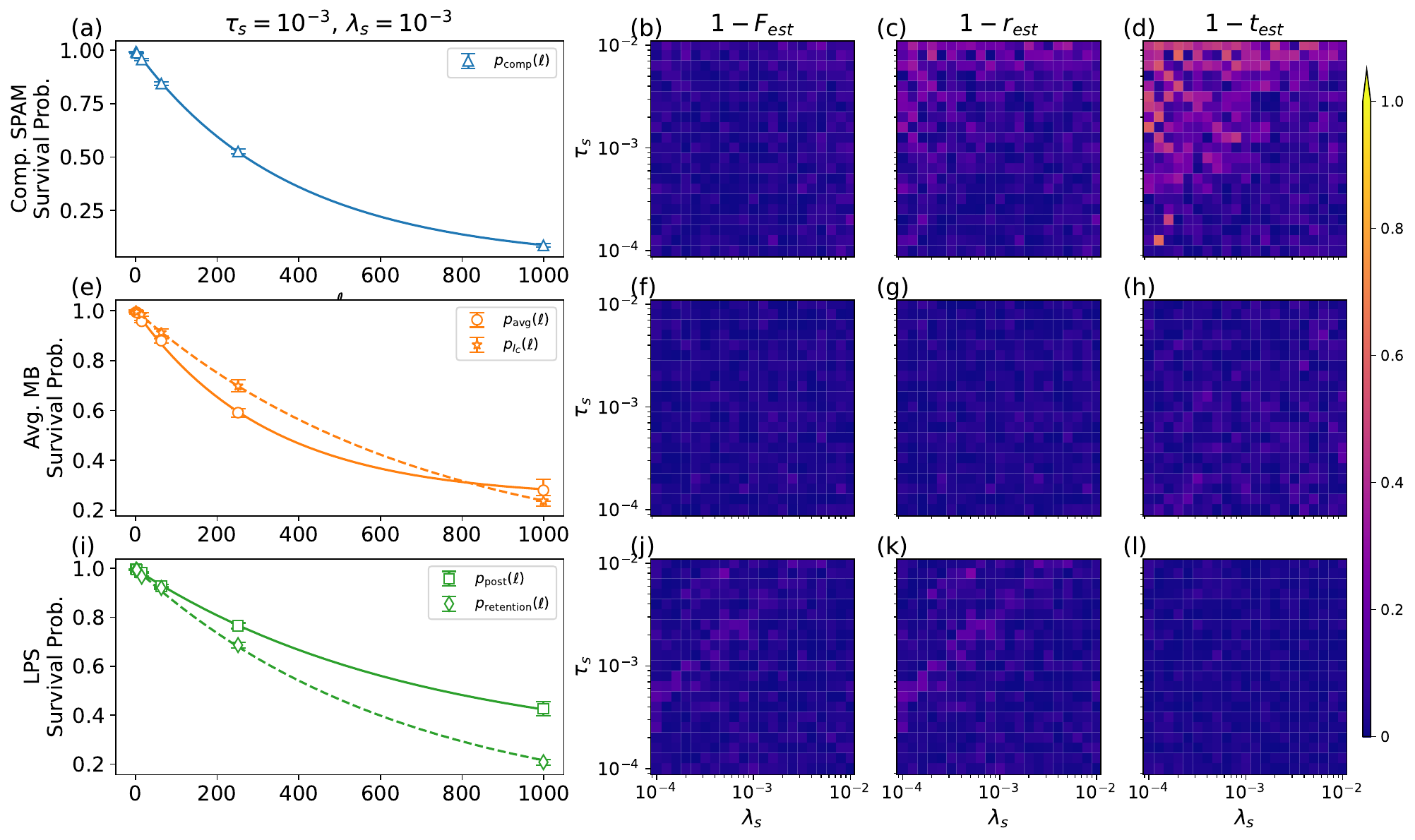}
\caption{Heat-map plots of the relative difference between the estimated values from each method compared to the values of the input error model using methods tailored to error processes without seepage (e.g. $\bar{\Lambda}_{CL} = 0$). The x-axis of each subplot shows the injected value of $\lambda_s$ (the magnitude of the computational error) and the y-axis shows the injected value of $\tau_s$ (the magnitude of the leakage error). (a) Example decay curve for $\lambda_s=\tau_s=10^{-3}$ for the Comp. SPAM method, (b) Infidelity $1-F$ for the Comp. SPAM method, (b) $1-r$ for the Comp. SPAM method, (b) $\tau$ for the Comp. SPAM method, (c) Example decay curve for $\lambda_s=\tau_s=10^{-3}$ for the Avg. MB method (d) infidelity $1-F$ for the Avg. MB method, (e) $1-r$ for the Avg. MB method, (f) $\tau$ for the Avg. MB method, (g) Example decay curve for $\lambda_s=\tau_s=10^{-3}$ for the LPS method (h) infidelity $1-F$ for the LPS method, (i) $1-r$ for the LPS method, and (j) $\tau$ for the LPS method.\label{fig:no_seepage}}
\end{figure*}

\subsubsection{Average over measurement basis}
Alternatively, one can use the Avg. MB method in Sec.~\ref{sec:rb_methods} to derive $r$ and $t$ separately. We choose a computational basis as the initial state $\rho_{\textrm{in}}=|i\>\<i|\in\chi_C$. Eq.~\eqref{eq:avg_basis_survival} gives
\begin{align}
    p_{\textrm{avg}}(\ell)&=\Tr\[\rho_{\textrm{in}}\bar{\Lambda}^{\ell}(\rho_\textrm{in})\]+\frac{1}{d_C}\Tr\[\mathds{I}_L\bar{\Lambda}^{\ell}(\rho_{\textrm{in}})\] \nonumber\\
    &=\frac{d_C-1}{d_C}r^{\ell}+\frac{1}{d_C},
\end{align}
where we use Eq.~\eqref{eq:2exp_fit} for $\Tr\[\rho_{\textrm{in}}\bar{\Lambda}^{\ell}(\rho_\textrm{in})\]$ and $\Tr\[\mathds{I}_L\bar{\Lambda}^{\ell}(\rho_{\textrm{in}})\]=1-\Tr\[\mathds{I}_C\bar{\Lambda}_{CC}^{\ell}(\rho_{\textrm{in}})\]=1-t^{\ell}$ from Eq.~\eqref{eq:lambda_CC_to_the_ell}. This gives a simple single exponential decay where the decay rate is $r$. To obtain $t$ (or the leakage rate $\tau$), we perform the computational identity measurement introduced in the Avg. MB method in Sec.~\ref{sec:rb_methods}. We use a measurement operator $|k\>\<k|$ that contains no leakage outcomes and set the initial state to be that projector $\rho_{\textrm{in}}=|k\>\<k|\in\chi_C$. From Eq.~\eqref{eq:comp_identity_survival}, we have
\begin{align}
    p_{\mathds{I}_C}(\ell)=\Tr\[\mathds{I}_C\bar{\Lambda}^{\ell}_{CC}(\rho_{\textrm{in}})\]=t^{\ell},
\end{align}
where the functional form from Eq.~\eqref{eq:lambda_CC_to_the_ell} is used. This gives a single exponential decay where the decay rate gives $t$ (and $\tau$). Combining with $r$, one can obtain the total fidelity $F$ from Eq.~\eqref{eq:fidelities0}.

\subsubsection{Leakage post-selection}\label{sec:lps_no_seep}
Finally, if one has the ability to post-select on the outcomes that contain no leakage as described in the LPS method in Sec.~\ref{sec:rb_methods}, then the post-selection data retention probability from Eq.~\eqref{eq:retention_survival} is 
\begin{align} \label{eq:no_seepage_retention}
p_{\textrm{retention}}(\ell)=\Tr\[\mathds{I}_C \bar{\Lambda}^{\ell}_{CC}\(\rho_{\textrm{in}}\)\]=t^{\ell},
\end{align}
where the functional form from Eq.~\eqref{eq:lambda_CC_to_the_ell} is used. Consider again $\rho_{\textrm{in}}=\Pi_{\textrm{out}}=|\psi\rangle\langle \psi|\in \chi_C$, the post-selected survival probability from Eq.~\eqref{eq:post_selected_survival} becomes
\begin{align}
p_{\textrm{post}}(\ell)&=\frac{p_{\textrm{comp}}(\ell)}{p_{\textrm{retention}}(\ell)}=\frac{d_C-1}{d_C}\(\frac{r}{t}\)^{\ell}+\frac{1}{d_C},
\end{align}
where the second equality uses Eq.~\eqref{eq:lambda_CC_to_the_ell}. Fitting this survival probability with a single exponential gives the ratio $r/t$.
Hence, together with $t$, we can deduce the fidelity $F$ from Eq.~\eqref{eq:fidelities0}.

\subsubsection{Simulations}

Results of a larger simulation for error process with a range of $\tau_s$ and $\lambda_s$ values are shown in  Fig.~\ref{fig:no_seepage}. In this case the modeled error process has no seepage. We see that the maximum relative difference between the estimated and the true infidelity is both 0.20 for the Comp. SPAM method, 0.12 for the Avg. MB method, and 0.20 for the LPS method. The leakage plot for Comp. SPAM, Fig.~\ref{fig:no_seepage}d, has larger disagreement when leakage errors dominate, $\lambda_s \ll \tau_s$ (upper left half). We believe this is due to the fitting instabilities of the double-exponential decay, which do not affect the infidelity estimate and could be alleviated with more sampling. 

The no seepage assumption is likely stronger than the previous two assumptions. This holds when the seepage rate is much smaller than the leakage error and the computational error, and the sequence length is chosen such that seepage is rare~\footnotemark[1]. The advantage is that the sequence length has no restriction with respect to the computational and leakage errors and the fits are simpler than the previous section. However, the LPS method still suffers from possibly low retention in long sequences, which may cause larger noise.

\subsection{Population transfer} \label{sec:pop_transfer}

Finally, we consider the twirled error process $\bar{\Lambda}$ that only moves population between computational and leakage states and maintain no phase information. Similar to the previous section, this regime makes no assumptions about relative error magnitudes but does require assumptions about the structure of the error process. In practice, these assumptions can be enforced with phase randomization~\cite{Epstein14, Wallman15, Wood18,Wu24} but that requires extra control of the system.

This regime requires an expanded treatment from what was shown in Sec.~\ref{sec:main_result}. Instead of considering the combined leakage subspace $\chi_L$, we decompose into separate leakage subspaces~\footnote[4]{In general, a system may have multiple leakage subspaces that have different population transfer rates. For example, two qubits have at least five different leakage subspaces due to the leakage subspaces of each individual qubit, e.g., $\{|0L\>,|1L\>,|L0\>,|L1\>,|LL\>\}$.}. We denote  $\{\mathds{I}_m\}_{m=0}^M$ the set of $M+1$ identity operators from each of the $M+1$ subspaces (one computational and $M-1$ leakage), where $\mathds{I}_0=\mathds{I}_C$ denotes the identity operator in the computational space. Our population transfer assumption implies that the only terms that are non-zero in $\bar{\Lambda}$ are $\Tr[P_{C,i} \bar{\Lambda}(P_{C,j})]$ and $\Tr[\mathds{I}_m \bar{\Lambda}(\mathds{I}_{m'})]$, where $P_{C,i}=\tfrac{1}{\sqrt{d_C}}P_i \oplus 0$ is a computational non-identity Pauli operator and there is zero projection on the leakage subspace. Therefore, $\bar{\Lambda}$ can be broken into two block diagonal operators with bases $\{P_{C,i}\}_{i=1}^{d_C^2-1}$ and $\{\mathds{I}_m\}_{m=0}^M$.

Recall that the twirled error process has the form
\begin{align}
    \bar{\Lambda}=\begin{pmatrix}
        r\mathcal{I}_C(\cdot)+(t-r)\Tr\[\mathcal{I}_C(\cdot)\]\tfrac{\mathds{I}_C}{d_C} & \bar{\Lambda}_{CL}\\
        \bar{\Lambda}_{LC} &\bar{\Lambda}_{LL}
    \end{pmatrix}.
\end{align}
We rewrite the upper-left block of the channel in the representation with $\Tr\[P_{C,i}\bar{\Lambda}(P_{C,j})\]$ for $i,j\neq 0$ and the part $\Tr\[\mathds{I}_C\bar{\Lambda}(\mathds{I}_C)\]$ explicitly separated out. Specifically, we write
\begin{align}
    \bar{\Lambda}=\begin{pmatrix}
        \begin{pmatrix}
            \bar{\Lambda}_C &0 \\
            0 & t\Tr\[\mathcal{I}_C(\cdot)\]\tfrac{\mathds{I}_C}{d_C}
        \end{pmatrix}& \bar{\Lambda}_{CL}\\
        \bar{\Lambda}_{LC} &\bar{\Lambda}_{LL}
    \end{pmatrix},
\end{align}
where $\bar{\Lambda}_C:=r\mathcal{I}_C(\cdot)-r\Tr\[\mathcal{I}_C(\cdot)\]\tfrac{\mathds{I}_C}{d_C}$ so that $\Tr\[\mathds{I}_C\bar{\Lambda}_C(\mathds{I}_C)\]=0$. The population transfer assumption implies that the only non-zero terms in $\bar{\Lambda}$ are $\bar{\Lambda}_C$ (computational errors) and identity terms in $\bar{\Lambda}_{LC}$, $\bar{\Lambda}_{CL}$, $\bar{\Lambda}_{LL}$ of the form $\bar{\Lambda}_{\mathds{I}}:=\Tr[\mathds{I}_m \bar{\Lambda}(\mathds{I}_{m'})]$. Therefore, $\bar{\Lambda}$ can be arranged as two block diagonal operators as
\begin{equation} \label{eq:symmetric_error}
    \bar{\Lambda} =\begin{pmatrix}
    \bar{\Lambda}_{C} & 0 & 0\\
    0 & \bar{\Lambda}_{\mathds{I}} & 0 \\
    0 & 0 & 0
\end{pmatrix}.
\end{equation}
The accumulated error process after a length $\ell$ sequence reads
\begin{align}
    \bar{\Lambda}^{\ell} =\begin{pmatrix}
    \bar{\Lambda}_{C}^{\ell} & 0 & 0\\
    0 & \bar{\Lambda}_{\mathds{I}}^{\ell} & 0 \\
    0 & 0 & 0
\end{pmatrix}, \label{eq:lambda_C_to_ell}
\end{align}
where $\bar{\Lambda}_C^{\ell}=r^{\ell}\mathcal{I}_C(\cdot)-r^{\ell}\Tr\[\mathcal{I}_C(\cdot)\]\tfrac{\mathds{I}_C}{d_C}$. 

In general, $\Lambda_{\mathds{I}}$ is a  $(M+1) \times (M+1)$ matrix, and therefore hard to diagonalize analytically for $M > 1$. The $M=1$ case is studied in Refs.~\cite{Wallman16, Claes21, Wood18}. Even if we had an explicit form for it, the eigenvalues may not relate to $t$ easily (Ref.~\cite{Wood18} resorts to using other fit parameters to differentiate $t$).

In order to derive survival probabilities for the Comp. SPAM and LPS methods we would need to make additional assumptions about $\bar{\Lambda}_{\mathds{I}}$ as is done in Ref.~\cite{Wu24}. Instead, we only consider the Avg. MB method, which can provide a tight bound on fidelity without additional assumptions. From Eq.~\eqref{eq:avg_basis_survival}, we have 
\begin{equation}\label{eq:pavg_survival}
    p_{\textrm{avg}}(\ell)=\Tr\[\rho_{\textrm{in}}\bar{\Lambda}^{\ell}(\rho_{\textrm{in}})\]+\frac{1}{d_C}\Tr\[\mathds{I}_L\bar{\Lambda}^{\ell}(\rho_{\textrm{in}})\].
\end{equation}
We can expand the computational initial state in the Pauli basis as $\rho_{\textrm{in}}=\mathds{I}_C/d_C+\sum_{i=1}^{d^2_C-1}v_iP_{C,i}$. Therefore, we have 
\begin{align}
    \Tr\[\rho_{\textrm{in}}\bar{\Lambda}^{\ell}(\rho_{\textrm{in}})\]=r^{\ell}\sum_{i=1}^{d^2_C-1}v^2_i+\frac{1}{d_C}\Tr\[\mathds{I}_C\bar{\Lambda}_{\mathds{I}}^{\ell}(\mathds{I}_C/d_C)\].
\end{align}
Now, we use the fact that $\rho_{\textrm{in}}$ is a projector, i.e., $\Tr\[\rho_{\textrm{in}}^2\]=1=\tfrac{1}{d_C}+\sum_{i=1}^{d^2_C-1}v^2_i$ which implies $\sum_{i=1}^{d^2_C-1}v^2_i=\tfrac{d_C-1}{d_C}$. Putting the above back to Eq.~\eqref{eq:pavg_survival} gives
\begin{align}
    p_{\textrm{avg}}(\ell)&=\frac{d_C-1}{d_C} r^{\ell}+\frac{1}{d_C}\Tr\[\bar{\Lambda}^{\ell}(\rho_{\textrm{in}})\]  \nonumber\\
    &=\frac{d_C-1}{d_C} r^{\ell}+\frac{1}{d_C},
\end{align}
where the second equality uses the trace-preserving property of $\bar{\Lambda}^{\ell}$. This gives a single exponential decay where the rate corresponds to $r$, which provides a bound on $F$ from Eq.~\eqref{eq:fidelity_bound}.

\subsubsection{Simulations}
\begin{figure} 
\centering
\includegraphics[width=\columnwidth]{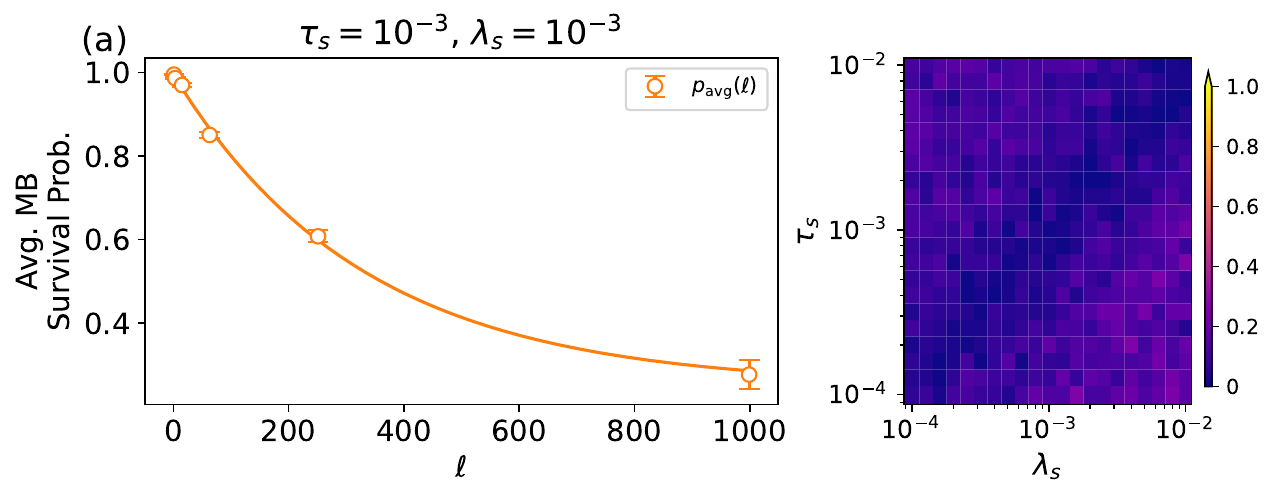}
\caption{Heat-map plot of the relative difference between the estimated values from each method compared to the values of the input error model using methods tailored to population transfer assumption. The x-axis shows the injected value of $\lambda_s$ (the magnitude of the computational error) and the y-axis shows the injected value of $\tau_s$ (the magnitude of the leakage error). (a) Example decay curve for $\lambda_s=\tau_s=10^{-3}$ for the Comp. SPAM method (b) $1-F$ average from the bound. 
\label{fig:pop_transfer}}
\end{figure}

Results of a larger simulation for error process with a range of $\tau_s$ and $\lambda_s$ values are shown in Fig.~\ref{fig:pop_transfer}. We chose to define the estimated infidelity as $\tfrac{7}{8} (1-r)$, which represents the midway position between the two bounds in Eq.~\eqref{eq:fidelity_bound}. We see that the maximum relative difference between the estimated and the true infidelity is 0.25, which is roughly one standard deviation plus the maximum distance 0.125 between the estimate and the bound.

The advantage in the leakage population transfer error regime is that it has no sequence lengths restrictions with respect to error parameters and is a single exponential decay which is easier to fit. The disadvantage is that it requires specific assumptions on the error process and cannot extract the leakage rate $\tau$ without even more assumptions.

\section{Comparison with Quantinuum system data}\label{sec:machine_data}
To demonstrate that leakage is having a measurable effect on current generation quantum computers infidelity estimates, we apply our analysis techniques to Quantinuum H1-1 and H2-1 2Q RB datasets from April 10, 2024 and May 20, 2024 in the repository Ref.~\cite{github_spec}. The previous estimate of infidelity was from a method similar to Avg. MB in the population transfer assumption but using the lower bound for infidelity $1-F= \tfrac{d_C-1}{d_C} (1-r)$~\cite{Pino21, Moses23}. Both datasets also included a leakage gadget at the end of every sequence that measured the leakage rate $\tau$ with a first order fitting, in the same method from LPS in Eq.~\eqref{eq:comp_dominant_retention_survival_prob}, but did not include that value in the infidelity estimate. For H1-1, the reported infidelity was $8.8(3)\times10^{-4}$ and the reported leakage rate was $2.1(3)\times10^{-4}$ and for H2-1 the reported infidelity was $1.36\times10^{-3}$ and the reported leakage rate was $3.3(4)\times10^{-4}$ (parentheses give one-sigma standard deviation from semi-parametric bootstrap resampling).

The Quantinuum datasets were taken before the development of these methods, and therefore do not match the error assumptions, sequence lengths, and circuit construction requirements for all methods. However, the datasets did include final basis permutations and a leakage gadget that enable some methods and also allow modifications to enable hybrid options. We made the following modifications to bring the data into the appropriate form for certain methods:
\begin{itemize}
    \item{Short sequence: We truncated the datasets to remove the longest two sequences for the H1-1 data and the longest sequence for the H2-1 data to apply methods for short sequences. This was necessary since the longest lengths violated the short sequence assumption.}
    \item{Comp. SPAM: Final randomization was used in the datasets so some sequences ended with measurement operators that contained projection into the leakage subspace. Specifically, the `1' output measurement measures both the $\ket{1}$ state population and the leakage state populations with state-dependent resonance florescence~\cite{Moses23}. To compensate, we used the leakage gadget to post-select on outcomes that did not contain leakage for these measurement projectors thereby recovering a projector only onto the computational subspace.}
    \item{Avg. MB: While the sequences had final measurement permutations the permutations were randomized so that the number of measurements of each basis state may not be equal for each sequence length. Instead of using the method from Eq.~\eqref{eq:comp_identity_survival} to measure the computational population we used the leakage gadget retention rate, as done in the LPS method, which also seems to offer more stability in the numerics from previous sections.}
\end{itemize}
We did not apply any analysis from Sec.~\ref{sec:no_seepage} since there is likely seepage in the experiment. We also did not apply the Comp. SPAM in the dominant computational error regime since we found empirically from simulations that the method requires long enough sequences to differentiate $\tau$ and $\lambda$.

\begin{figure} 
\centering
\includegraphics[width=\columnwidth]{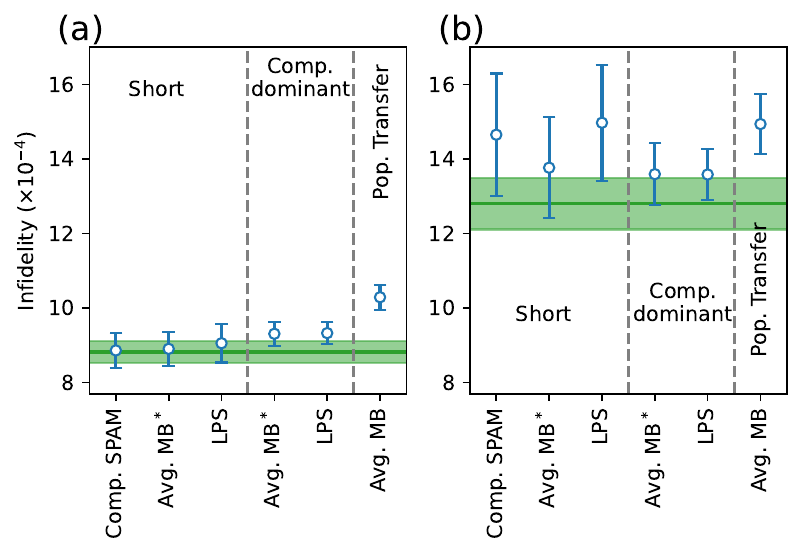}
\caption{Analysis of Quantinuum systems from data shared in Ref.~\cite{github_spec}. Each plot shows the estimated 2Q infidelity with different methods (blue circles) compared to the previously reported 2Q infidelity (green line). The data is for the following systems: (a) H1-1 with five gate zones and 20 qubits taken on April 10, 2024 and (b) H2-1 with four gate zones and 56 qubits taken on May 20, 2024. \label{fig:h_series_data} }
\end{figure}

The results of the new analyses are plotted as blue markers in Fig.~\ref{fig:h_series_data} and the previous infidelity estimate is plotted as a green line. Most updated methods that include leakage show a greater than one standard deviation increase in the estimated infidelity. The main exception is the short sequence linear fitting methods for H1-1, which also had larger errors bars, since some data had to be removed, and are still within roughly one-sigma of other methods. The Avg. MB for the population transfer returned the largest estimate for both the H1-1 data and H2-1 data. This is likely because that estimate picks the middle point between the two bounds from Eq.~\eqref{eq:fidelity_bound} and from the leakage gadget retention measurements it is clear that the computational errors are dominating, which makes the Avg. MB estimate overly conservative. 

We believe the best estimates are from the Avg. MB and LPS methods with the computational dominant error regime. All methods agree that $\lambda > \tau$ and $\tau \ell \ll 1$, which is the computational dominant error regime. We saw from the simulations in Fig.~\ref{fig:comp_dominant} that the estimate of $r$ (or $\lambda$) performed well for both of these methods and the leakage gadget retention estimated $\tau$ well (which is what is used for both Avg. MB and LPS here). Both methods are also consistent with each other: H1-1 Avg. MB $9.3(3)\times10^{-4}$ and LPS $9.3(3)\times10^{-4}$, and H2-1 Avg. MB $1.36(8)\times10^{-3}$ and LPS $1.36(7)\times10^{-3}$.

\section{Conclusions}
The presence of leakage errors invalidates many common RB practices. This is especially true when leakage errors are a substantial fraction of the total errors in the system and often cause RB to underestimate total infidelity. We showed several methods for properly accounting for leakage errors in RB in different, realistic error regimes. We derived these methods with a new and simplified approach for RB with leakage and verified the performance with numerical simulation of 2Q RB. We also reanalyzed previously shared data from Quantinuum's H1-1 and H2-1 systems and showed that leakage, while small, has a non-negligible contribution to infidelity. As systems improve, our analysis will be vital to RB methods since leakage errors often represent fundamental limits on gates that will begin to dominate.

Our analysis only focused on standard Clifford RB but all versions of RB likely must also account for leakage. Many of the methods we derived will translate to other RB variants but will require more work in the future to verify they are properly accounting for leakage errors. Moreover, there are other potential affects in RB that we have not considered like other non-Markovian errors that are known to cause drift in standard RB~\cite{Epstein14,Fogarty15,Fong17}. We believe our methods could be compatible with different approaches to include these effects in RB decay fitting but leave it to future work.

As quantum computers continue to improve performance many platforms will begin to move towards QEC, where leakage is particularly harmful~\cite{Brown20}. Many proposals for correcting or mitigating leakage will need to be verified with techniques like RB. Two such examples are: (1) leakage repumping that moves leaked population to computational spaces deterministically~\cite{Hayes20}, and (2) physical leakage detection that identifies a qubit that has been leaked~\cite{Kang23}. Method 1, leakage repumping, eliminates the need for special RB protocols if it removes leakage completely. It also may reduce the total infidelity since leakage errors move all states to orthogonal states but computational errors only move a fraction to orthogonal states. Method 2, leakage detection, is similar to the gadget we described earlier but has smaller error rates that allow for fault-tolerant QEC. For either method it will be important to use RB, or other tools, to verify that the methods do not add any new computational errors and properly benchmarking performance and estimate fidelity to understand the effect in QEC.

\begin{acknowledgments}
We thank the entire Quantinuum team for contributions to the datasets in Ref.~\cite{github_spec} and feedback on leakage benchmarking. We especially thank Karl Mayer, Michael Mills, and Justin Gerber for helpful discussions.
\end{acknowledgments}

\appendix

\section{Survival probabilities with SPAM error}\label{app:SPAM_surv} 
Here, we discuss the effect of SPAM errors on survival probabilities. Suppose $\Lambda_P$ is the state preparation error process and $\Lambda_M$ is the measurement error process, which also includes the error of the final inversion gate in the RB sequence. 
In general, a survival probability decay including SPAM error is
\begin{align}
    p_{\textrm{SPAM}}(\ell)=\Tr\[\Pi_{\textrm{out}}\Lambda_M\bar{\Lambda}^{\ell}\Lambda_P(\rho_{\textrm{in}})\].
\end{align}
We can rewrite the noisy initial state as $\Lambda_P(\rho_{\textrm{in}})=\rho_{\textrm{in}}+\delta\rho$ and the noisy measurement operator as $\Pi_{\textrm{out}}\Lambda_M=\Pi_{\textrm{out}}+\delta\Pi$, where $\rho_{\textrm{in}}$ and $\Pi_{\textrm{out}}$ represent the ideal initial state and measurement projector and $\delta\rho$ and $\delta\Pi$ represent the errors induced by $\Lambda_P$ and $\Lambda_M$. The survival probability can be written as
\begin{align}
    &p_{\textrm{SPAM}}(\ell) \nonumber\\
    &=p(\ell)+\Tr\[\Pi_{\textrm{out}}\bar{\Lambda}^{\ell}(\delta\rho)\]+\Tr\[\delta\Pi\bar{\Lambda}^{\ell}(\rho_{\textrm{in}})\] \nonumber\\
    &\ \ \ \ \ \ \ \ \ \ +\Tr\[\delta\Pi\bar{\Lambda}^{\ell}(\delta\rho)\],
\end{align}
which implies
\begin{align}
    |p_{\textrm{SPAM}}(\ell)-p(\ell)|=\OO(\delta\rho)+\OO(\delta\Pi).
\end{align}
This means the survival probability including SPAM error only deviates from the ideal decay at the order of the state-preparation and measurement error. The explicit effect of SPAM error can vary for different error regimes and fitting methods as we elaborate below.

As a first example, in the short sequences regime for the computational SPAM method in Sec.~\ref{sec:small}, the accumulated error process is expanded up to the first-order in  error components, i.e., $\bar{\Lambda}^{\ell}\approx\mathcal{I}+\ell\mathcal{E}$. The survival probability of preparing $\rho_{\textrm{in}}=\Pi_{\textrm{out}}\in\chi_C$ is
\begin{align} \label{eq:first_order_long0}
    p_{\textrm{SPAM}}(\ell) &= \Tr\[ \Pi_{\textrm{out}}\Lambda_M \bar{\Lambda}^{\ell}\Lambda_P(\rho_{\textrm{in}})\] \nonumber\\
    &\approx \Tr\[ \Pi_{\textrm{out}}\Lambda_M (\mathcal{I}+\ell\mathcal{E})\Lambda_P(\rho_{\textrm{in}})\] \nonumber \\
    &= A - \ell\[\lambda(B-C)+\tau B-D\]
\end{align}
where 
\begin{align}
    A &= 1+\Tr\[\delta\Pi\rho_{\textrm{in}}\] + \Tr\[\Pi_{\textrm{out}}\delta\rho\] + \Tr\[\delta\Pi\delta\rho\], \nonumber \\
    B &=1+\Tr_C\[\delta\Pi\rho_{\textrm{in}}\] + \Tr_C\[\Pi_{\textrm{out}}\delta\rho\] + \Tr_C\[\delta\Pi\delta\rho\],\ \nonumber \\
    C &= \tfrac{1}{d_C}\left(1+\Tr_C\[\delta\rho\]+\Tr_C\[\delta\Pi\]+\Tr_C\[\delta\rho\]\Tr_C\[\delta\Pi\]\right), \nonumber \\
    D &= \Tr\[\delta\Pi(\bar{\Lambda}_{CL}+\bar{\Lambda}_{LC})(\rho_{\textrm{in}})\] \nonumber \\
    &+\Tr\[\Pi_{\textrm{out}}(\bar{\Lambda}_{CL}+\bar{\Lambda}_{LC})(\delta\rho)\]  \nonumber \\
    &+\Tr\[\delta\Pi\mathcal{E}_{LL}(\delta\rho)\],
\end{align}
where $\Tr_C \[ \cdot\]:=\Tr\[\mathcal{I}_C(\cdot)\]$ is the trace over the computational subspace. For small SPAM errors, i.e., $|\delta\rho|\ll1$ and $|\delta \Pi|\ll1$, we have $A\approx1$, $B\approx1$, $C\approx1/d_C$ and $D\approx0$. SPAM errors will cause an underestimation of infidelity since $B$, $C$, and $D$ all enter the slope in Eq.~\eqref{eq:first_order_long0}. However, the sizes of these errors are $\mathcal{O}(\delta\Pi)+\mathcal{O}(\delta\rho)$. A similar underestimation of infidelity from SPAM errors is also expected in the linear decay fit of the computational dominant regime in Sec.~\ref{sec:comp_dominant}. For these cases, all SPAM error terms directly enter and affect the rates of the fit functions.

On the other hand, for exponential decays, some SPAM error terms can be separated from the error rates in the fits. As an example, we consider the no seepage case (Sec.~\ref{sec:no_seepage}), where the computational-to-computational block is
\begin{align}
\(\bar{\Lambda}^{\ell}\)_{CC}=\bar{\Lambda}^{\ell}_{CC}=r^{\ell}\mathcal{I}_C+\(t^{\ell}-r^{\ell}\)\Tr\[\mathcal{I}_C(\cdot)\]\mathds{I}_C/d_C
\end{align}
from Eq.~\eqref{eq:lambda_CC_to_the_ell}. The survival probability including SPAM becomes
\begin{align}
    p_{\textrm{SPAM}}(\ell)&=\Tr\[ \Pi_{\textrm{out}}\Lambda_M \bar{\Lambda}^{\ell}\Lambda_P(\rho_{\textrm{in}})\] \nonumber\\
    &=Ar^{\ell}+Bt^{\ell}+C(\ell),
\end{align}
where 
\begin{align}
    A&=\frac{d_C-1}{d_C}+\Tr_C\[\Pi_{\textrm{out}}\delta\rho\]-\Tr_C\[\delta\rho\]/d_C \nonumber\\
    &+\Tr_C\[\delta\Pi\rho_{\textrm{in}}\]-\Tr_C\[\delta\Pi\]/d_C \nonumber\\
    &+\Tr_C\[\delta\Pi\delta\rho\]-\Tr_C\[\delta\Pi\]\Tr_C\[\delta\rho\]/d_C, \nonumber\\
    B&=\frac{1}{d_C}\left(1+\Tr_C\[\delta\rho\]+\Tr\[\delta\Pi\]+\Tr\[\delta\Pi\]\Tr\[\delta\rho\]\right)
\end{align}
and $C(\ell)$ is a sequence length dependent SPAM error term of size $\mathcal{O}(\delta\Pi)+\mathcal{O}(\delta\rho)$, from the accumulative effect of $\bar{\Lambda}_{LC}$, $\bar{\Lambda}_{CL}$, and $\bar{\Lambda}_{LL}$. The SPAM error terms entering $A$ and $B$ do not affect the estimates of rates $r$ and $t$ but the terms from $C(\ell)$ might. A similar relation holds for the exponential decays from the computational dominant error regime in Sec.~\ref{sec:comp_dominant} and the population transfer in Sec.~\ref{sec:pop_transfer}.

Overall, we expect the contribution of SPAM errors to estimates to be small for all the methods and cases provided in Sec.~\ref{sec:survival_probs}, when the SPAM error magnitudes are small. In the numerical simulation we included measurement error and showed that the estimates were generally close outside of specific areas discussed in Sec.~\ref{sec:survival_probs}.

\section{Block-diagonal form of an ideal gate}\label{app:block-diagonal_form}
Suppose $C$ is a noiseless gate acting on the joint Hilbert space $\HH=\HH_C\oplus\HH_L$. The corresponding unitary can be written as a 2-by-2 block form, i.e.,
\begin{align}
C=\begin{pmatrix}
    C_C & C_{CL} \\
    C_{LC} & C_L
\end{pmatrix}
\end{align}
where $C_C \in \chi_C$, $C_L \in\chi_L$, and $C_{CL}$ and $C_{LC}$ represent transformations between $\HH_C$ and $\HH_L$. In practice, we wish to implement $C_C$ on the computational space for quantum computations but the unitary on the leakage subspace $C_L$ is often uncontrolled and the result of unintended operators that are system specific. 

Since an ideal gate does not have leakage error, it implies $C_{LC}=0$. In addition, $C$ should be a unitary on the total space, which implies
\begin{align}
&C^{\dagger}C=\begin{pmatrix}
    C^{\dagger}_CC_C & C^{\dagger}_CC_{CL} \\
    C^{\dagger}_{CL}C_C & C^{\dagger}_{CL}C_{CL}+C^{\dagger}_LC_L
\end{pmatrix}=\begin{pmatrix}
    I_C & 0\\
    0 & I_L
\end{pmatrix} \\
&\implies C_{CL}=0,\ C^{\dagger}_CC_C=I_C,\ \text{and}\  C^{\dagger}_LC_L=I_L.
\end{align}
This shows $C$ has to be in the form of
\begin{align}
    C=\begin{pmatrix}
        C_C & 0 \\
        0 & C_L
    \end{pmatrix},
\end{align}
where $C_C$ and $C_L$ are unitaries in the respected spaces. The corresponding ideal process $\C$ also follows this block-diagonal form.

\section{The leakage identity approximation}\label{app:leak_identity_approx}
In the Avg. MB method of Sec.~\ref{sec:rb_methods}, we make an approximation that
\begin{align}
    \sum_k\Tr\[Q^{\dagger}_k\Pi_{L,k}Q_k\bar{\Lambda}^{\ell}(\rho_{\textrm{in}})\]\approx \Tr\[\mathds{I}_{L}\bar{\Lambda}^{\ell}(\rho_{\textrm{in}})\].
\end{align} 
For 1Q leakage, this amounts to assuming $Q_k$ has no action on the leakage states, $Q_k^{\dagger}\Pi_{L,k}Q_k=\Pi_{L,k}$. This is likely not true for more than one qubits since one such leakage state is a single qubit leaks but other qubits do not leak. Since $Q_k$ is built from 1Q $I,X$ gates then some gates will alter the state of that leakage subspace.

\begin{figure} 
\centering
\includegraphics[width=\columnwidth]{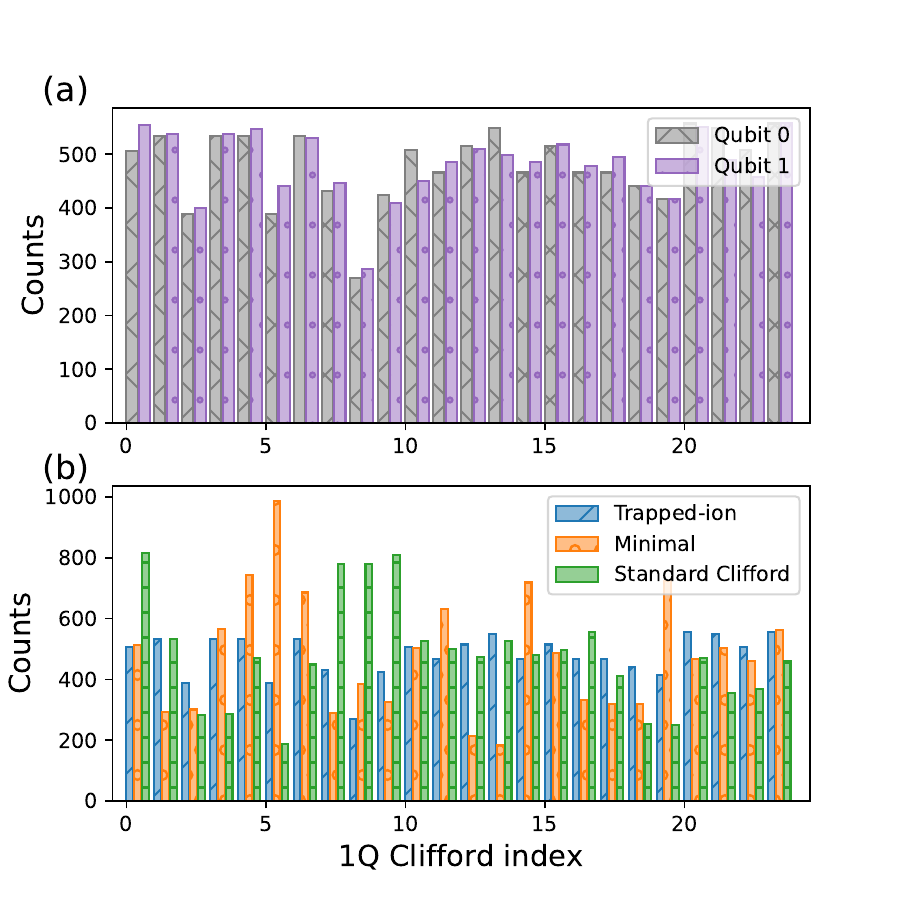}
\caption{Distribution of 1Q gates across the one qubit leaked subspaces. (a) Comparison of the qubit 0 and qubit 1 distribution for a selected gateset composed of $R_{ZZ}(\pi/2)$ and $\pm \pi/2$ and $\pi$ rotations around $X, Y$, and $Z$ axes. (b) Comparison of distribution from the previous gateset in (a) called ``Trapped-ion'' to a minimal gateset with $CNOT$ and $+\pi/2$ rotations around $X$ and $Y$ and a standard Clifford generating set with $CNOT$, $H$, and $P$.  \label{fig:sq_clifford_distribution} }
\end{figure}

To treat two or more qubits, we analyze the approximate action of the random gates when one or more qubits are leaked. For simplicity, we consider two qubits but it is possible to generalize to more qubits. First, write the identity in the leakage subspace
\begin{align}
    &\mathds{I}_L\\
    &=|0L\>\<0L|+|1L\>\<1L|+|L0\>\<L0|+|L1\>\<L1|+|LL\>\<LL|,\nonumber
\end{align}
which assumes a single leakge state $|L\>$ but could be replaced with a summation over other states. In general, each leakage projection operator $\Pi_{L,k}$ can have some component in each leakage state, i.e.,
\begin{align}
    \Pi_{L,k}=\sum_{x\in\{0L,1L,L0,L1,LL\}} c_{k,x}|x\>\<x|.
\end{align}
Again, we make the assumption that single-qubit gates do not affect non-computational states, i.e., $X|L\>=|L\>$. In Fig.~\ref{fig:sq_clifford_distribution} we see empirically the 2Q Cliffords generated from a few typical gatesets are approximately uniform 1Q Cliffords when one qubit is leaked and the two-qubit gates do not interact with the remaining qubit. An approximately random 1Q Clifford then randomizes the unleaked qubit such that
\begin{align}
 &\<0L|\bar{\Lambda}^{\ell}(\rho_{\textrm{in}})|0L\>\approx\<1L|\bar{\Lambda}^{\ell}(\rho_{\textrm{in}})|1L\>, \nonumber\\
 &\<L0|\bar{\Lambda}^{\ell}(\rho_{\textrm{in}})|L0\>\approx\<L1|\bar{\Lambda}^{\ell}(\rho_{\textrm{in}})|L1\>. \label{eq:unleaked_sq_randomized}
\end{align}
The POVM property $\sum_{k}\Pi_{L,k}=\mathds{I}_L$, implies $\sum_{k}c_{k,x}=1$ for all $x$. There are four ways that we case permute the computational basis, i.e., $Q_{II}=II,\ Q_{IX}=IX,\ Q_{XI}=XI$ and $Q_{XX}=XX$. We find that either $Q_k|x\>\<x|Q_k=|x\>\<x|$ (i.e., when $Q_k$ acts on $L$ part) or $\Tr\[Q_k|x\>\<x|Q_k\bar{\Lambda}^{\ell}(\rho_{\textrm{in}})\]=\Tr\[|x\>\<x|\bar{\Lambda}^{\ell}(\rho_{\textrm{in}})\]$ due to Eq.~\eqref{eq:unleaked_sq_randomized}. So we have
\begin{align}
    &\sum_{k\in\{II,IX,XI,XX\}}\Tr\[Q_{k}\Pi_{L,k}Q_{k}\bar{\Lambda}^{\ell}(\rho_{\textrm{in}})\] \nonumber\\
    &=\sum_{k}\sum_{x}c_{k,x}\Tr\[Q_{k}|x\>\<x|Q_k\bar{\Lambda}^{\ell}(\rho_{\textrm{in}})\] \nonumber\\
    &=\sum_{k}\sum_{x}c_{k,x}\Tr\[|x\>\<x|\bar{\Lambda}^{\ell}(\rho_{\textrm{in}})\] \nonumber\\
    &=\sum_{x}\Tr\[|x\>\<x|\bar{\Lambda}^{\ell}(\rho_{\textrm{in}})\]=\Tr\[\mathds{I}_L\bar{\Lambda}^{\ell}(\rho_{\textrm{in}})\].
\end{align}
This completes the argument. 

\section{Identity for $\bar{\Lambda}_{CC}^{\ell}$}\label{app:proof_for_lambda_to_l}
Given a trace-non-preserving depolarizing process
\begin{align}
    \Lambda(\rho)=a\mathcal{I}(\rho)+ b \Tr\left[\mathcal{I}(\rho)\right]\frac{\mathds{I}}{d},
\end{align}
for arbitrary $a$ and $b$. We prove the following identity 
\begin{align}
\Lambda^{\ell}(\cdot)=a^{\ell}\mathcal{I}(\cdot) + \left[(a+b)^{\ell}-a^{\ell}\right]\Tr\left[\mathcal{I}(\cdot)\right]\frac{\mathds{I}}{d}
\end{align}
for any values $a$ and $b$.

First observe that 
\begin{align}
&\Lambda^2\left(\rho\right)=\Lambda\left(\Lambda\left(\rho\right)\right) \nonumber\\
&=a\left(a\rho+b \Tr\left[\rho\right]\frac{\mathds{I}}{d}\right) + b \Tr\left[a\rho+b \Tr\left[\rho\right]\frac{\mathds{I}}{d}\right]\frac{\mathds{I}}{d} \nonumber\\
&=a^2\rho +(ab+ba+b^2)\Tr\left[\rho\right]\frac{\mathds{I}}{d} \nonumber\\
&=a^2\rho+\left[(a+b)^2-a^2\right]\Tr[\rho]\frac{\mathds{I}}{d}.
\end{align}
Suppose the following formula holds for $\ell$ (which is true for $\ell=1,2$),
\begin{align}
\Lambda^{\ell}(\rho)=a^{\ell}\rho+\left[(a+b)^{\ell}-a^{\ell}\right]\Tr\left[\rho\right]\frac{\mathds{I}}{d}.
\end{align}
Then
\begin{align}
\Lambda^{\ell+1}(\rho)&=a\left(a^{\ell}\rho+\left[(a+b)^{\ell}-a^{\ell}\right]\Tr\left[\rho\right]\frac{\mathds{I}}{d}\right)\nonumber \\
&\ \ \ \ +b \Tr\left[a^{\ell}\rho+\left[(a+b)^{\ell}-a^{\ell}\right]\Tr\left[\rho\right]\frac{\mathds{I}}{d}\right]\frac{\mathds{I}}{d} \nonumber\\
&=a^{\ell+1}\rho +a \left[(a+b)^{\ell}-a^{\ell}\right]\Tr\left[\rho\right]\frac{\mathds{I}}{d}\nonumber\\
&\ \ \ \ +\left[ba^{\ell}+b(a+b)^{\ell}-ba^{\ell}\right]\Tr[\rho]\frac{\mathds{I}}{d} \nonumber\\
&=a^{\ell+1}\rho+ \left[(a+b)^{\ell+1}-a^{\ell+1}\right]\Tr\left[\rho\right]\frac{\mathds{I}}{d}
\end{align}
holds for $\ell+1$. Therefore, by induction, we proved
\begin{align}
&\Lambda^{\ell}(\rho)=a^{\ell}\rho+\left[(a+b)^{\ell}-a^{\ell}\right]\Tr\left[\rho\right]\frac{\mathds{I}}{d} \nonumber\\
&\implies \Lambda^{\ell}(\cdot)=a^{\ell}\mathcal{I}(\cdot) + \left[(a+b)^{\ell}-a^{\ell}\right]\Tr\left[\mathcal{I}(\cdot)\right]\frac{\mathds{I}}{d}
\end{align}
for any positive integer $\ell$. Now we substitute $a=r$ and $b=t-r$, we get
\begin{align}
\bar{\Lambda}^{\ell}_{CC}(\rho)=r^{\ell}\mathcal{I}_C(\rho) + \left(t^{\ell}-r^{\ell}\right)\Tr\left[\mathcal{I}_C(\rho)\right]\mathds{I}_C/d_C.
\end{align} 

\section{Derivations for Sec.~\ref{sec:comp_dominant}}\label{app:comp_large_eq}
Here, we provide detailed derivations for the survival probabilities given in Sec.~\ref{sec:comp_dominant}. Let us define the depolarizing process as
\begin{align}
    \Lambda_{\textrm{dep}}(\cdot):=(1-\lambda)\mathcal{I}_C+\lambda\Tr\[\mathcal{I}_C(\cdot)\](\mathds{I}_C/d_C).
\end{align}
It is shown in App. \ref{app:proof_for_lambda_to_l} that 
\begin{align}
    \Lambda_{\textrm{dep}}^{\ell}(\cdot):=(1-\lambda)^{\ell}\mathcal{I}_C+\[1-(1-\lambda)^{\ell}\]\Tr\[\mathcal{I}_C(\cdot)\](\mathds{I}_C/d_C).
\end{align}
Recall from Eq.~\eqref{eq:comp_dominant_expansion}, we have
\begin{align}
    \Lambda_{\textrm{comp}}=\begin{pmatrix}
        \Lambda_{\textrm{dep}} &0\\
        0& \mathcal{I}_L
    \end{pmatrix}
\end{align}
and 
\begin{align}
    \Lambda_{\textrm{comp}}^{\ell}=\begin{pmatrix}
        \Lambda_{\textrm{dep}}^{\ell} &0\\
        0& \mathcal{I}_L
    \end{pmatrix}
\end{align}
for any positive integer $\ell$. Also from Eq.~\eqref{eq:comp_dominant_expansion}, we have
\begin{align}
    \mathcal{E}'=\begin{pmatrix}
        -\tau\mathcal{I}_C &\bar{\Lambda}_{CL} \\
        \bar{\Lambda}_{LC} & \mathcal{E}_{LL}
    \end{pmatrix}.
\end{align}
Simple algebra gives
\begin{align}
&\Lambda_{\textrm{comp}}^{m}\mathcal{E}'\Lambda_{\textrm{comp}}^{n} \nonumber\\
&=\begin{pmatrix}
    -\tau\Lambda_{\textrm{dep}}^{m+n} & \Lambda_{\textrm{dep}}^{m}\bar{\Lambda}_{CL} \\
   \bar{\Lambda}_{LC}\Lambda_{\textrm{dep}}^{n}  & \mathcal{E}_{LL}
\end{pmatrix},
\end{align}
for any positive integers $m$ and $n$. We have
\begin{align}\label{eq_app:lambda_to_ell}
\bar{\Lambda}^{\ell}&=\(\Lambda_{\textrm{comp}}+\mathcal{E}'\)^{\ell} \approx \Lambda_{\textrm{comp}}^{\ell}+ \sum_{k=1}^{\ell}\Lambda_{\textrm{comp}}^{\ell-k}\mathcal{E}'\Lambda_{\textrm{comp}}^{k-1}\nonumber\\
&=\begin{pmatrix}
        \Lambda^{\ell}_{\textrm{dep}}-\ell \tau\Lambda^{\ell-1}_{\textrm{dep}}& \sum_{k=1}^{\ell} \Lambda_{\textrm{dep}}^{\ell-k}\bar{\Lambda}_{CL}\\
        \sum_{k=1}^{\ell}\bar{\Lambda}_{LC}\Lambda_{\textrm{dep}}^{k-1} & \mathcal{I}_L + \ell\mathcal{E}_{LL}
    \end{pmatrix}.    
\end{align}
Preparing $\rho_{\textrm{in}}=\Pi_{\textrm{out}}=|k\>\<k|\in \chi_C$, we have
\begin{align}\label{eq_app:exp_lin_decay}
    &p(\ell)=\Tr\[\Pi_{\textrm{out}}\bar{\Lambda}^{\ell}(\rho_{\textrm{in}})\] \nonumber\\
    &\approx \frac{d_C-1}{d_C}(1-\lambda-\ell \tau)(1-\lambda)^{\ell-1}+(1-\ell \tau)\frac{1}{d_C}.
\end{align}
This gives the survival probability in Eq.~\eqref{eq:comp_dominant_survival_prob}.

For the method of averaging over measurement basis, we have
\begin{align}
    p_{\textrm{avg}}(\ell)=\Tr\[\rho_{\textrm{in}}\bar{\Lambda}^{\ell}(\rho_\textrm{in})\]+\frac{1}{d_C}\Tr\[\mathds{I}_L\bar{\Lambda}^{\ell}(\rho_{\textrm{in}})\],
\end{align}
where the first term is given by Eq.~\eqref{eq_app:exp_lin_decay} and the second term is 
\begin{align}
    &\frac{1}{d_C}\Tr\[\mathds{I}_L\bar{\Lambda}^{\ell}(\rho_{\textrm{in}})\]=\frac{1-\Tr\[\mathds{I}_C\bar{\Lambda}^{\ell}(\rho_{\textrm{in}})\]}{d_C} \nonumber\\
    &\approx\frac{1-\Tr\[\Lambda_{\textrm{dep}}^{\ell}(\rho_{\textrm{in}})\]+\ell\tau\Tr\[\Lambda^{\ell-1}_{\textrm{dep}}(\rho_{\textrm{in}})\]}{d_C}=\frac{\ell\tau}{d_C}, \label{eq:linear_leakage_decay}
\end{align}
where we use the upper-left block of Eq.~\eqref{eq_app:lambda_to_ell} for the second equation. Therefore, we have 
\begin{align}
    p_{\textrm{avg}}(\ell)\approx\frac{d_C-1}{d_C}(1-\lambda-\ell\tau)(1-\lambda)^{\ell}+\frac{1}{d_C}.
\end{align}
Note that in this error regime, we choose sequence length $\ell$ such that $\ell\tau\ll1$. One can use a binomial expansion in terms of $1-\lambda$ and $-\tau$ for $(1-\lambda-\tau)^{\ell}$, i.e.,
\begin{align}
    &(1-\lambda-\tau)^{\ell}=\sum_{k=0}^{\ell} {\ell \choose k}(1-\lambda)^{\ell-k}(-\tau)^k\nonumber\\
    &=(1-\lambda)^{\ell}+\ell(-\tau)(1-\lambda)^{\ell-1}+\OO(\ell^2\tau^2) \nonumber\\
    &=(1-\lambda-\ell\tau)(1-\tau)^{\ell-1}+\OO(\ell^2\tau^2).
\end{align}
This implies
\begin{align}
    p_{\textrm{avg}}(\ell)\approx\frac{d_C-1}{d_C}(1-\lambda-\tau)^{\ell}+\frac{1}{d_C}+\OO(\ell^2\tau^2).
\end{align}
Eq.~\eqref{eq:linear_leakage_decay} above also implies that the computational identity measurement in Eq.~\eqref{eq:comp_dom_identity_meas} is
\begin{align}
    p_{\mathds{I}_C}(\ell)=\Tr\[\mathds{I}_C\bar{\Lambda}^{\ell}(|k\>\<k|)\]\approx 1-\ell\tau. \label{eq_app:comp_identiy_leak_surv}
\end{align}

\begin{table*}[]
\begin{tabular}{|l|c|c|c|}
\hline
                                                                                                                 & Comp. SPAM & Avg. MB & LPS \\ \hline \hline
\begin{tabular}[c]{@{}l@{}}Small lengths\\ Sec.~\ref{sec:small}\end{tabular}               & $\textrm{Space}\[1, \tfrac{1}{25}\textrm{min}(\tfrac{1}{\lambda_s},\tfrac{1}{\tau_s}), 6\]$  & $\textrm{Space}\[1,\tfrac{1}{25}\textrm{min}(\tfrac{1}{\lambda_s},\tfrac{1}{\tau_s}), 6\]$ &$\textrm{Space}\[1, \tfrac{1}{25}\textrm{min}(\tfrac{1}{\lambda_s},\tfrac{1}{\tau_s}), 6\]$\\ \hline
\begin{tabular}[c]{@{}l@{}}Comp. dominant\\ Sec.~\ref{sec:comp_dominant}\end{tabular}     & $\textrm{Space}\[1, \textrm{max}(\tfrac{1}{\lambda_s},\tfrac{1}{25 \tau_s}), 6\]$ & $10^{\textrm{Space}\[0, -\log{\lambda_s}, 6\]}$ & $10^{\textrm{Space}\[0, -\log\lambda_s,  6\]}$\\ \hline
\begin{tabular}[c]{@{}l@{}}No seepage\\ Sec.~\ref{sec:no_seepage}\end{tabular}            & $10^{\textrm{Space}\[0, -\log\[\min(\lambda_s,\tau_s)\],  6\]}$ &$10^{\textrm{Space}\[0, -\log\[\min(\lambda_s,\tau_s)\],  6\]}$ & $10^{\textrm{Space}\[0, -\log\[\min(\lambda_s,\tau_s)\],  6\]}$\\ \hline
\begin{tabular}[c]{@{}l@{}}Population transfer\\ Sec.~\ref{sec:pop_transfer}\end{tabular} & - & $10^{\textrm{Space}\[0, -\log\[\min(\lambda_s,\tau_s)\],  6\]}$ & -\\ \hline
\end{tabular}
\caption{The RB sequence lengths used in each simulations. Space[$x, y, z$] returns $z$ evenly spaced numbers (rounded to the nearest integer) between $x$ and $y$\label{tab:RB_lengths}}
\end{table*}

In the leakage post-selection setting, one is able to determine whether the system is in the leakage subspace at the end of the shot and only keep those that remain in the computational space. With post-selection, one can always set the measurement and the initial state as computational basis state projectors $\rho_{\textrm{in}}=\Pi_{\textrm{out}}=|k\rangle\langle k|$. The post-selected survival probability is then
\begin{align}
    p_{\textrm{post}}(\ell)&=\frac{\Tr\[|k\rangle\langle k|\bar{\Lambda}^{\ell}(|k\rangle\langle k|)\]}{\Tr\[\mathds{I}_C\bar{\Lambda}^{\ell}(|k\rangle\langle k|)\]}=\frac{\Tr\[|k\rangle\langle k|\bar{\Lambda}^{\ell}(|k\rangle\langle k|)\]}{p_{\textrm{retention}}(\ell)} \nonumber\\
    &\approx\frac{d_C-1}{d_C}\frac{(1-\lambda-\ell \tau)}{1-\ell \tau}(1-\lambda)^{\ell-1}+\frac{1}{d_C}, 
\end{align}
where the $\Tr\[|k\rangle\langle k|\bar{\Lambda}^{\ell}(|k\rangle\langle k|)\]$ is given by Eq.~\eqref{eq_app:exp_lin_decay} and the data retention probability is $p_{\textrm{retention}}(\ell)=\Tr\[\mathds{I}_C\bar{\Lambda}^{\ell}(|k\>\<k|)\]\approx 1-\ell \tau$ using Eq.~\eqref{eq_app:comp_identiy_leak_surv}. If one further makes the approximation to drop any term of second order, i.e., terms with $\mathcal{O}(\ell \tau \lambda)$ and higher, then we have
\begin{align}
p_{\textrm{post}}(\ell)\approx\frac{d_C-1}{d_C}(1-\lambda)^{\ell}+\frac{1}{d_C},
\end{align}
which is a single exponential decay with the decay corresponding to the computational error $\lambda$.

\section{Simulation details} \label{app:numerics}
The leakage process chosen for simulations incoherently connects both qubit levels to the leaked state $\ket{l}$ modeled by a Lindblad master equation with leakage jump operators $\ketbra{l}{1}$ and $\ketbra{l}{0}$, seepage jump operators $\ketbra{1}{l}$ and $\ketbra{0}{l}$, and scattering probability $\gamma$,
\begin{align} \label{eq:example_process}
    \dot{\rho} &= \gamma \sum_{i=0}^1\ketbra{l}{i} \rho \ketbra{i}{l} - \tfrac{1}{2}\left(\ketbra{i}{i} \rho + \rho \ketbra{i}{i}\right) \nonumber \\
    &+ \frac{\gamma}{2} \sum_{i=0}^1\ketbra{i}{l} \rho \ketbra{l}{i} - \tfrac{1}{2}\left(\ketbra{i}{i} \rho + \rho \ketbra{i}{i}\right), 
\end{align}
where the first line is for leakage and the second line is for seepage. For small times, we can approximate the process as $\Lambda_L(\rho) = \rho - \Delta t \dot{\rho}$ for time step $\Delta t$. The process act symmetrically on each qubit with the same magnitude $\gamma$ for total process $\Lambda^{\otimes 2}_L$. The magnitude in the main text is $\tau_s = \delta t \gamma$. This process has fidelity $F=1-\tau_s$, depolarizing parameter $r=1-\tau_s$, and computational population $t=1-\tau_s$. For this error regime, we model a similar process without the seepage jump operators (remove second line from Eq.~\eqref{eq:example_process}) and it has the same fidelity, depolarizing parameter, and computational population.

We also include a depolarizing process that acts only on the computational subspace
\begin{equation}
    \Lambda_{\textrm{dep}}(\rho) = (1-\lambda_s) \mathcal{I}_{C}(\rho)+\tfrac{\lambda_s}{d_C}\Tr\[\mathcal{I}_C(\rho)\]\mathds{I}_C,
\end{equation}
like in Eq.~\ref{eq:comp_depolarizing} but trace preserving on the computational subspace. This process has fidelity $F=1-\tfrac{3}{4}\lambda_s$, depolarizing parameter $r=1-\lambda_s$, and computational population $t=1$.

The total error is then $\Lambda = \Lambda_{\textrm{dep}} + \Lambda^{\otimes 2}_L$ with fidelity $F=1-\tfrac{3}{4}\lambda_s-\tau_s$, depolarizing parameter $r=1-\lambda_s-\tau_s$, and computational population $t=1-\tau_s$.

We also include SPAM errors to better model an experiment. The first contribution is a measurement error that independently bit flips each qubit with probability $\lambda_s$. For LPS methods, we add a second contribution to SPAM errors of two extra error processes at the end of each circuit to mimic the error from a leakage detection gadget. We do not include any 1Q gate errors.

For each RB method, we select sequence lengths with the rules in Table~\ref{tab:RB_lengths}.

\bibliography{references}

\end{document}